\documentclass[11pt,letterpaper]{article}
\usepackage{fullpage,indentfirst,rotating}
\usepackage{epsfig}
\usepackage{graphicx}
\usepackage{rotating}
\usepackage{subfigure}
\usepackage{color}
\usepackage{epsfig}
\usepackage{graphicx}
\usepackage{rotating}
\usepackage{subfigure}
\usepackage{amsfonts}
\usepackage{latexsym}
\usepackage{amsmath}
\usepackage{amssymb}
\usepackage{amsthm}
\usepackage{amstext}
\usepackage[colorlinks=true, linkcolor=blue, citecolor=blue]{hyperref}
\usepackage{soul}
\usepackage[authoryear]{natbib}
\usepackage{todonotes}
\usepackage{comment}
\usepackage{setspace}
\usepackage{multirow}
\usepackage[flushleft]{threeparttable}
\allowdisplaybreaks

\newcommand{\bt}{\beta}

\newcommand{\dt}{\delta}
\newcommand{\ld}{\lambda}

\newcommand{\tht}{\theta}
\newcommand{\Tht}{\Theta}

\newcommand{\rhot}{\rho_{\tau}}
\newcommand{\eps}[0]{\ensuremath{\varepsilon}}

\newcommand{\lt}{\left}
\newcommand{\rt}{\right}
\newcommand{\mk}{_{(m,k)}}
\newcommand{\imk}{_{i(m,k)}}

\newcommand{\ave}{\frac{1}{n}\sum_{i=1}^n}
\newcommand{\sumi}{\sum_{i=1}^n}
\newcommand{\summ}{\sum_{m=1}^{M}}
\newcommand{\sumk}{\sum_{k=1}^{K}}
\newcommand{\maxi}{\max_{1 \le i \le n}}
\newcommand{\maxk}{\max_{1 \le k \le K}}
\newcommand{\maxm}{\max_{1 \le m \le M}}
\newcommand{\avem}{M^{-1}\sum_{m=1}^{M}}

\newcommand{\ben}{\begin{enumerate}}
\newcommand{\een}{\end{enumerate}}
\newcommand{\bit}{\begin{itemize}}
\newcommand{\eit}{\end{itemize}}

\newcommand{\rarrow}[0]{\ensuremath{\rightarrow}}

\newcommand{\darrow}[0]{\ensuremath{\stackrel{d}{\rightarrow}}}

\newcommand{\eq}[1]{\begin{align}#1\end{align}}
\newcommand{\eqs}[1]{\begin{align*}#1\end{align*}}
\newcommand{\what}[0]{\ensuremath{\widehat}}
\newcommand{\wtd}[0]{\ensuremath{\widetilde}}

\newcommand{\changed}[1]{#1}

\DeclareMathOperator*{\argmax}{arg\,max}
\DeclareMathOperator*{\argmin}{arg\,min}

\newtheorem{assum}{Assumption}

\newtheorem{theorem}{Theorem}
\newtheorem{lemma}{Lemma}

\usepackage[linesnumbered,ruled]{algorithm2e}
\SetKwInput{KwInput}{Input}                
\SetKwInput{KwOutput}{Output}              

\input{tcilatex}
\begin{document}

\title{Complete Subset Averaging for Quantile Regressions
\thanks{The authors would like to thank \changed{the Editor, Peter Phillips, the Co-Editor, Arthur Lewbel, }and three anonymous referees for helpful comments and suggestions, which \changed{have} led to substantial improvements. They also thank Xun Lu and Liangjun Su for helpful discussion and sharing their codes. Shin is grateful \changed{for partial} support by
 the Social Sciences and Humanities Research Council of Canada (SSHRC-435-2018-0275). This work was made possible by the facilities of WestGrid (www.westgrid.ca) and Compute Canada (www.computecanada.ca).}}
\author{ Ji Hyung Lee\thanks{
Department of Economics, University of Illinois, 214 David Knley Hall, 1407 West Gregory Drive, Urbana, IL 61801, USA, jihyung@illinois.edu.} \and %
Youngki Shin\thanks{
Department of Economics, McMaster University, 1280 Main St. W., Hamilton, ON L8S 4L8, Canada, shiny11@mcmaster.ca.}}
\date{\today}

\maketitle

\begin{abstract}
We propose a novel conditional quantile prediction method based \changed{on complete} subset averaging (CSA) for quantile regressions. All models under consideration are potentially misspecified and the dimension of regressors goes to infinity as the sample size increases. Since we average over the complete subsets, the number of models is much larger than the usual model averaging method which adopts sophisticated weighting schemes. We propose to use an equal weight but select the proper size of the complete subset based on the leave-one-out cross-validation method. Building upon the theory of \citet{lu2015jackknife}, we investigate the large sample properties of CSA and show the asymptotic optimality in the sense of \citet{li1987asymptotic}. We check the finite sample performance via Monte Carlo simulations and empirical applications.
\medskip

\noindent \textit{Keywords: }\ complete subset averaging, quantile regression, prediction, equal-weight, model
averaging.

\vspace{0.08in}

\noindent \textit{\ JEL classification: }C21, C52, C53

\newpage
\end{abstract}

\section{Introduction}

Quantile regression (QR) has emerged as an essential tool since \cite{koenker1978regression} (see, e.g.\ \citet{koenker_2005}). QR estimates the response of conditional quantiles of outcome
variables with respect to changes in the covariates. The entire response
distribution of outcome variables in economic models provides a broader
insight than the classical mean regression. Moreover, in many economic
applications, \changed{tail quantiles} have highly valuable information. See, for
example, wage distribution in labor economic applications \citep{buchinsky1998dynamics} and stock return quantiles
(Value-at-Risk) in financial market analysis \citep{duffie1997overview}.
Recently, policymakers have begun to pay attention to the left tail quantiles of
GDP growth (Growth-at-Risk)
as a measure of downside risks associated with tight financial conditions \citep*{adrian2019vulnerable}.
There has also been an increasing interest in climate change, in particular, more frequent and intense extreme weather conditions. A tail quantile is the main object of interest in this analysis \citep*{bhatia2019recent}. Estimation, inference, and
prediction of the conditional quantiles are thus important but require a
careful econometric analysis due to their nonlinear structure and
nonstandard limit theory.

In this paper, we propose a novel prediction method based \changed{on complete} subset averaging (CSA) for quantile regressions. Following \citet{lu2015jackknife}, we work on the framework such that all models under consideration are potentially misspecified and that the dimension of regressors goes to infinity as the sample size increases. The CSA method that we propose works as follows. First, pick the numbers of regressors $k$ out of all regressors $K$ available in the data. Then, there exist $K!/(k!(K-k)!)$ complete subsets of size $k$. Second, estimate all the quantile regression models and save all the conditional quantile predictors from each model. Finally, the conditional quantile predictor is constructed as the average of all the quantile predictors estimated in Step 2.
Since we average over the complete subsets, the number of models is much larger than the usual model averaging methods selecting the weight of each model. We propose to use an equal weight but select the optimal size of the complete subset $k^*$ based on the leave-one-out cross-validation method.

The CSA approach has a couple of advantages over the existing model averaging method which adopts sophisticated weighting schemes. First, it may produce better forecasts in practice because there is no sampling variance from the weight estimation. This result is already reported both in the forecasting and machine learning literature in the mean regression setup (see, e.g.\ \citet{breiman1996bagging}, \citet{clemen1989combining}, \citet{stock2004combination}, \citet{smith2009simple}, and \citet*{elliott2013complete}). Second, it does not ask a researcher to choose the initial set of models and the order of each model. In practice, the model averaging methods with different weights usually construct the set of models in an encompassing way and the forecasting performance could depend on the researcher's discretion. Third, CSA averages over a larger number of submodels and one could expect an additional noise reduction from it. However, CSA is possibly more demanding in computation, and we will discuss this issue in detail later.

The contribution of this paper is twofold. First, building upon the theory of \citet{lu2015jackknife}, we show that the complete subset quantile regression (CSQR) estimator converges the pseudo-true value and satisfies asymptotic normality under mild regularity conditions. The uniform convergence property of CSQR is also provided. Based on these pointwise and uniform limit theories, we prove the asymptotic optimality of $\what{k}$ in the sense of \citet{li1987asymptotic}. Second, we implement the CSA method and show that it performs quite well both in simulations and real data sets. Especially, we show that the performance is still satisfactory when we use a fixed number of subsets randomly drawn from the complete subsets when the time budget does not allow estimating the quantile regressions of the whole subsets. We also provide regularity conditions on the choice of the fixed number of subsets. Finally, we provide a theory that compares the performance of equal weighting and optimal weighting in quantile regression. This result justifies our intuition such that optimal weighting forecasts poorly when the number of models increases and extends the existing result in mean regression.


Finally, we summarize related literature. \citet{lu2015jackknife} and \citet*{elliott2013complete} are closely related to this paper. The former proposes the jackknife model averaging (JMA) method for the quantile prediction problem and derives the nonstandard asymptotic properties of the estimator. Our approach is different from theirs in  that we use complete subsets for models to be averaged and that we choose a \emph{scalar} $\what{k}$ from the cross-validation method instead of a weighting vector $\what{w}$. The latter proposes the CSA method in the mean prediction problem and shows by simulation studies that the CSA predictor outperforms alternative methods like bagging, ridge, lasso, and Bayesian model averaging. However, they do not show any optimality result of the estimator. \citet{hansen2007least} and \citet{hansen2012jackknife} show the optimality of model averaging based on the Mallows criterion and  that of the jackknife model averaging, respectively. \citet{ando2014model} propose a model averaging method in a high-dimensional setting and show the optimality result. \citet{komunjer2013quantile} provides a great review on the quantile prediction problem of time-series data. \citet{meinshausen2006quantile} proposes a quantile prediction method based on random forest. \citet{lee2016predictive} studies the inference problem of the predictive quantile regression when the regressors are persistent.
In the empirical finance literature, \citet*{meligkotsidou2019quantile,meligkotsidou2021out} apply complete subset quantile regression to forecast realized volatility and the risk premium.

The rest of the paper is organized as follows. Section \ref{sc: model} introduces the
model and the CSQR estimator. Section \ref{sc: asympt} presents the asymptotic properties of the CSQR estimator and the asymptotic optimality. The Monte Carlo simulation results are reported in Section \ref{sc:simulations}. Section \ref{sc:empirical} investigates two empirical applications and illustrates the advantage of the proposed method. Section \ref{sc:concl} concludes.
\changed{All the proofs are deferred to the appendix.}

We use the following notation. For a matrix $A$, $\left\Vert \cdot
\right\Vert $ represents its Frobenius norm $\left\Vert A\right\Vert =\sqrt{tr(AA^{\prime })}$.
Let $\lambda _{\min }\left( A\right) $ and $\lambda _{\max
}\left( A\right) $ denote the smallest and largest eigenvalues of $A$.
We use the notation $x_n \approx y_n$ to denote $x_n=y_n+o_p(1)$; and $a_n \ll b_n$ to denote $a_n=o(b_n)$.

\bigskip

\section{Model and Estimator}\label{sc: model}

In this section, we lay out the model under study and propose the complete
subset averaging (CSA) quantile predictor. We also discuss the
choice of the subset size based on the cross-validation method.

\subsection{CSA Quantile Predictor}

Consider a random sample $\{(y_{i},x_{i}^{\prime })\}$ for $i=1,\ldots ,n$,
where the dimension of $x_{i}$ can be countably infinite. Following \citet{lu2015jackknife},
we assume that $\{(y_{i},x_{i}^{\prime })\}_{i=1}^n$ is generated from  the following
linear quantile regression model: for $\tau \in (0,1)$,
\begin{equation}
y_{i}=\mu _{i}+\varepsilon _{i}=\sum_{j=1}^{\infty }\theta
_{j}x_{ij}+\varepsilon _{i},  \label{eq-dgp1}
\end{equation}
where $\mu _{i}=\mu _{i}(\tau ):=\sum_{j=1}^{\infty }\theta _{j}x_{ij}$, $%
\theta _{j}=\theta _{j}(\tau )$, $\varepsilon _{i}=\varepsilon _{i}(\tau
):=y_{i}-Q_{y}(\tau |x_{i})$, and $Q_y(\tau|x)$ is the $\tau$-th conditional quantile function of $y$ given $x$.  Note that we drop $\tau$ from each expression for notational simplicity and that $\varepsilon _{i}$ satisfies the
quantile restriction $P(\varepsilon _{i}(\tau )\leq 0|x_{i})=\tau $. Equivalently, we can also express $Q_{y}\left( \tau |x_{i}\right)
:=\sum_{j=1}^{\infty }\theta _{j}(\tau )x_{ij}$ as is often done in the
quantile regression literature. 

We consider a sequence of covariates available, which approximate the above
quantile regression model:
\begin{equation*}
y_{i}=\sum_{j=1}^{K_n}\theta _{j}(\tau )x_{ij}+b_{i}(\tau )+\varepsilon
_{i}(\tau ),
\end{equation*}%
where $b_i=b_i(\tau):=\mu_i(\tau) - \sum_{j=1}^{K_n}\theta _{j}(\tau )x_{ij}$ is the approximation error
and $K_n$ is the total number of available regressors that may increase as the sample
size $n$ increases. Thus, we presume that all models are misspecified in a finite sample as in \citet{hansen2007least}.\footnote{\changed{Using quantile crossings, \citet{phillips2015halbert} also shows that quantile regresssion is always at the risk of model misspecification unless the parameters are local to constant over $\tau$.} } 

Given $K_n$ regressors, we consider a model composed of $k$ regressors, where $k \in \{1,2,\ldots,K_n\}$. There are $\frac{K_n!}{k!(K_n-k)!}$
different ways to select $k$ regressors out of $K_n$.
Therefore, a subset of size $k$ is composed of $M_{(K_n,k)}=\frac{K_n!}{k!(K_n-k)!}$ different elements and a model is defined as a single element of them. We use index $m_{(K_n,k)} \in \{1,2, \ldots ,M_{(K_n,k)}\}$ for each model.
For example, consider that we have $K_n=3$ regressors $\{x_{i1}, x_{i2}, x_{i3}\}$ and construct a subset of size $k=2$. Then, we have $M_{(3,2)}=3$ different ways to choose a model as follows: $(x_{i1}, x_{i2}), (x_{i1}, x_{i3}),$ and $(x_{i2}, x_{i3})$. Each model is indexed by $m_{(3,2)}\in\{1,2,3\}$.
For succinct notation, we drop all
subscripts from $K_n$, $M_{(K_n,k)}$, and $m_{(K_n,k)}$ and denote them as $K$, $M$, and $m$ unless there is any confusion.

We now consider a quantile regression model with regressors in a complete subset. Let model $m$ with a size $k$ be given. For observation $i$, let $x_{i(m,k)}$ be a $k$-dimensional vector of regressors corresponding to model $m$, i.e.\ $x_{i(2,2)}=(x_{i1},x_{i3})$ in the above example. We can construct a linear quantile regression model with regressors $x_{i(m,k)}$:
\eq{
y_{i}=x_{i(m,k)}^{\prime }\Theta _{(m,k)}+b_{i(m,k)}+\varepsilon _{i},\label{eq:main-quantile}
}
where $b_{i(m,k)}:=\mu _{i}-x_{i(m,k)}^{\prime }\Theta _{(m,k)}$ is again the approximation error when we use only $x_{i(m,k)}$ regressors. The model \changed{in} \eqref{eq:main-quantile} is estimated by the standard method in linear quantile
regression:
\begin{align}
\widehat{\Theta }_{(m,k)} & =\argmin_{\Theta _{(m,k)} \in \mathbf{\Tht}}\sum_{i=1}^{n}\rho _{\tau
}\left( y_{i}-x_{i(m,k)}^{\prime }\Theta _{(m,k)}\right) \label{eq:Tht_mk}\\
& := \argmin_{\Theta _{(m,k)} \in \mathbf{\Tht}} Q_{n}\left( \Theta _{(m,k)}\right)
\end{align}
where $\mathbf{\Tht}$ is a parameter space and $\rho _{\tau }(u):=u(\tau -1\{u\leq 0\})$ is the check function. Note that the estimator $\what{\Tht}_{(m,k)}$ is defined for each subset size $k$ and for each model $m$ with $k$ regressors. As noted above, we can think of $M$ different models and corresponding estimators that have $k$ regressors.

We have a few remarks here. First, we use the subscript $(m,k)$ to denote a generic model with $k$ regressors. However, the index set $\{1,\ldots,M_{(K_n,k)}\}$ itself is defined in terms of $k$, which implies that $m$ is also determined by $k$. Recall the original notation $m_{(K_n,k)}$ above. Therefore, model $m \in \{1,\ldots,M_{(K_n,k)}\}$ has the same number of regressors $k$ and we cannot choose $m$ and $k$ in an arbitrary way.
Second, we allow that the subset size $k$ goes to infinity as $n$ increases. In other words, there exists a sequence of subset sizes $\{k(n)\}$ that diverges. This setting is natural as the upper bound $K_n$ goes to infinity as $n$ increases. Note that the number of regressors in each model ($k_m$ in their notation) is also allowed to diverge in \citet{lu2015jackknife}\changed{. Both approaches} allow more complex models to be averaged as $n$ grows, which is measured by $k$ and $k_m$, respectively. However, \citet{lu2015jackknife} require controlling the growth rates of $M$ and $\max_m k_m$, separately. The proposed method constructs submodels based on the complete subsets, and $M$ is tightly related to $K$ and $k$. As a result, the regularity condition on the complexity of the models is expressed only in terms of $K_n$ (see Assumption \ref{a-3} in Section \ref{sc: asympt}).

We finalize this subsection by defining the complete subset averaging (CSA) quantile predictor. Let the size of the complete subset $k$ be given. For each model, we estimate the parameter $\what{\Tht}_{(m,k)}$ by \eqref{eq:Tht_mk} and construct the linear index $x_{(m,k)}'\what{\Tht}_{(m,k)}$. The CSA quantile predictor of $y$ given $x$ is defined as a simple average of those indices over $M$ different models:
\begin{equation*}
\widehat{y}(k)=\frac{1}{M}\sum_{m=1}^{M}x_{(m,k)}^{\prime }\widehat{\Theta }_{(m,k)}.
\end{equation*}
The CSA quantile predictor is different from the JMA quantile predictor of \citet{lu2015jackknife} in two respects. First, we do not select the set of models to be averaged since we average over the complete subsets of size $k$. Second, CSA does not estimate the weights over different models. The idea of averaging over the complete subsets was first introduced by \citet*{elliott2013complete} in the conditional mean prediction setup. Heuristically speaking, since the weights can be seen as additional parameters to be estimated in the model, the equal weight could perform better in a finite sample when the number of models (i.e.\ the dimension of a weight vector) is large.

\subsection{Choice of Subset Size \texorpdfstring{$k$}{Lg}}
We propose to choose the subset size $k$ using the leave-one-out cross-validation method. We will show in the next section that the subset size $\what{k}$ chosen by this method is optimal in the sense that it is asymptotically equivalent to the infeasible optimal choice.

For $k=1,\ldots,K$, we define a cross-validation objective function as follows:
\begin{align}
CV_n(k) & = \frac{1}{n}\sum_{i=1}^n \rho_{\tau} \left(y_i - \frac{1}{M}%
\sum_{m=1}^{M} x_{i(m,k)}^{\prime }\widehat {\Theta}_{i(m,k)}\right) \\
& = \frac{1}{n}\sum_{i=1}^n \rho_{\tau} \left(y_i - \widehat {y}_i(k)\right),
\end{align}
where $\widehat {\Theta}_{i(m,k)}$ is the jackknife estimator for $\widehat {%
\Theta}_{(m,k)}$, which is estimated by \eqref{eq:Tht_mk} without using the $%
i $-th observation $(x_i,y_i)$, and $\widehat {y}_{i}(k)$ is a corresponding jackknife
CSA quantile predictor for the $i$-th outcome variable $y_i$. The prediction error is measured by the check function $\rho_{\tau}(\cdot)$.
Then, we can choose the
complete subset size $k$ that minimizes the cross-validation objective
function as follows:
\begin{align}
\widehat {k} = \argmin_{1 \le k \le K} = CV_n(k).\label{eq:khat}
\end{align}
After choosing the complete subset size, the CSA quantile predictor is finally defined as
\eq{
	\what{y}(\what{k}) = \frac{1}{M} \summ x'_{(m,\what{k})} \what{\Tht}_{(m,\what{k})}\changed{,}
}where the plugged-in $\what{k}$ is chosen by \eqref{eq:khat}.

We finalize this subsection by adding some remarks on computation. First, we propose to use a fixed number $M_{max}$ of random draws of models when $M$ is too large to implement the method. Since $M=K!/(k!(K-k)!)$, it can be quite large when the model has large potential regressors. The simulation studies in Section \ref{sc:simulations} reveal that the CSA quantile predictor still performs well with a feasible size of submodels randomly drawn from the complete subsets. We also provide regularity conditions that assure the asymptotic equivalence between using $M$ and $M_{max}$ in Section \ref{sc: asympt}. Second, the proposed jackknife method can be immediately extended to the $b$-fold cross-validation method, where $b$ is the partition size of the sample.  Algorithm \ref{algo:CV} below summarizes the leave-one-out cross-validation method for choosing $\what{k}$.

\begin{algorithm}[hp]
 \KwInput{$\{(y_i, x_i): i=1,\ldots,n \}$, $M_{max}$}
\KwOutput{$ \widehat{k}$}
Set $K = dim(x_i)$\;
\For{$k=1$ \KwTo $K$}{
	Set $\mathcal{X}_{i,k} = \{\mbox{all combinations with $k$ regressors out of $x_i$}\}$\;
	Set $M = \vert \mathcal{X}_{i,k} \vert_0 = K!/(k!(K-k)!) $\;
	\If{$M \le M_{max}$}{
		\For{$m=1$ \KwTo $M$}{
		Set $x_{i(m,k)} = (\mbox{the $m$-th element of $\mathcal{X}_{i,k}$})$ for $i=1,\ldots,n$\;
			\For{$i=1$ \KwTo $n$}{
				Estimate the jackknife estimator $\widehat{\Theta }_{i(m,k)}$:
				\begin{equation}
					\widehat{\Theta }_{i(m,k)} =\argmin_{\Theta _{(m,k)} \in \mathbf{\Tht}}\sum_{j=1, j\neq i}^{n}\rho _{\tau
					}\left( y_{j}-x_{j(m,k)}^{\prime }\Theta _{(m,k)}\right) \label{eq:jackknife}
				\end{equation}
			}
		}
		Set $\what{y}_i(k) = \frac{1}{M}\summ x_{i(m,k)}'\what{\Theta}_{i(m,k)}$\;
		Set $CV_n(k) = \ave \rho_{\tau}(y_i - \what{y}_i(k))$\;
	}
	\If{$M > M_{max}$}{
		\For{$m=1$ \KwTo $M_{max}$}{
			Set $x_{i(m,k)} = (\mbox{a random element of $\mathcal{X}_{i,k}$})$ for $i=1,\ldots,n$\;
			\For{$i=1$ \KwTo $n$}{
				Estimate the jackknife estimator $\widehat{\Theta }_{i(m,k)}$ using \eqref{eq:jackknife}\;
			}
		}
		Set $\what{y}_i(k) = \frac{1}{M}\summ x_{i(m,k)}'\what{\Theta}_{i(m,k)}$\;
		Set $CV_n(k) = \ave \rho_{\tau}(y_i - \what{y}_i(k))$\;
	}
}
Set $\what{k}=\argmin_{k} CV_n(k)$\;
 \caption{Cross-validation for CSA}\label{algo:CV}
\end{algorithm}


\section{Asymptotic Theory}\label{sc: asympt}
In this section, we investigate the asymptotic properties of the complete subset quantile regression (CSQR) estimator. We first provide the pointwise and uniform convergence results of $\widehat{\Tht}_{(m,k)}$ and $\widehat{\Tht}_{i(m,k)}$, respectively. Then, we show the optimality of CSA in the sense of \citet{li1987asymptotic}, which implies that $\widehat{k}$ is asymptotically equivalent to the infeasible optimal choice of the subset size.

In addition to the model described in Section \ref{sc: model}, we define some notation for later use. Let $f_{y|x}(\cdot|x)$ be a conditional probability density function for generic random variables $x$ and $y$. Since all models are potentially misspecified in the model averaging literature, we define the pseudo-true parameter value for any given $(m,k)$:
\begin{equation*}
\Theta _{(m,k)}^{\ast }:=\argmin_{\Theta _{(m,k)} \in \mathbf{\Tht}}E\left[ \rho _{\tau
}\left( y_{i}-x_{i(m,k)}^{\prime }\Theta _{(m,k)}\right) \right].
\end{equation*}
Let $\psi_{\tau}(c) :=\tau - 1\{ c \le 0 \}$. For any $(m,k)$ such that $m=1,\ldots,M$ and $k=1,\ldots,K$, we define
\begin{eqnarray*}
A_{(m,k)} &:=&E\left[ f_{y|x}\left( \Theta _{(m,k)}^{\ast \prime
}x_{i(m,k)}|x_{i}\right) x_{i(m,k)}x_{i(m,k)}^{\prime }\right],\\
B_{(m,k)} &:=&E\left[ \psi _{\tau }\left( y_{i}- \Theta _{(m,k)}^{\ast \prime
}x_{i(m,k)}\right) ^{2}x_{i(m,k)}x_{i(m,k)}^{\prime }\right],
\end{eqnarray*}%
and
\begin{equation*}
V_{(m,k)}:=A_{(m,k)}^{-1}B_{(m,k)}A_{(m,k)}^{-1}\text{.}
\end{equation*}


We need the following regularity conditions.
\begin{assum}
\label{a-1}

\begin{enumerate}
\item[(i)] $(y_{i},x_{i})$ is i.i.d.\ generated by (\ref{eq-dgp1})

\item[(ii)] $P(\varepsilon _i(\tau) \le 0 |x_i) = \tau$ a.s.\

\item[(iii)] $E[\mu_i^4] < \infty$ and $\sup_{j \ge 1} E[x_{ij}^8] < c_x$
for some $c_x < \infty$
\end{enumerate}
\end{assum}

\begin{assum}
\label{a-2}

\begin{enumerate}
\item[(i)] $f_{y|x}(\cdot|x_i)$ is bounded above by $c_f<\infty$ and
continuous over its support a.s.

\item[(ii)] There exist constants $\underline{c}_{A(m,k)}$ and $\overline{c}%
_{A(m,k)}$ such that $0<\underline{c}_{A(m,k)} \le
\lambda_{\min}\left(A_{(m,k)}\right) \le
\lambda_{\max}\left(A_{(m,k)}\right) \le c_f \lambda_{\max}\left(E\left[%
x_{i(m,k)} x_{i(m,k)}^{\prime }\right]\right) \le \overline{c}_{A{(m,k)}}
<\infty$

\item[(iii)] There exist constants $\underline{c}_{B(m,k)}$ and $\overline{c}%
_{B(m,k)}$ such that $0< \underline{c}_{B(m,k)} \le \lambda_{\min} \left(
B_{(m,k)} \right) \le \lambda_{\max} \left( B_{(m,k)} \right) \le \overline{c%
}_{B(m,k)} < \infty$

\item[(iv)] $\left(\overline{c}_{A(m,k)} + \overline{c}_{B(m,k)}\right) / k
= O\left(\underline{c}_{A(m,k)}^2\right)$
\end{enumerate}
\end{assum}

\begin{assum}
\label{a-3} Let $\underline{c}_A := \min_{1 \le k \le K} \min_{1 \le m \le
M} \underline{c}_{A(m,k)}$, $\underline{c}_B := \min_{1 \le k \le K} \min_{1
\le m \le M} \underline{c}_{B(m,k)}$, $\overline{c}_A := \max_{1 \le k \le
K} \max_{1 \le m \le M} \overline{c}_{A(m,k)}$, and $\overline{c}_B :=
\max_{1 \le k \le K} \max_{1 \le m \le M} \overline{c}_{A(m,k)}$.

\begin{enumerate}
\item[(i)] $\frac{K^4\overline{c}_A}{n\underline{c}%
_B} =o(1)$ and $\frac{K^4 (\log n)^4}{n \underline{c}_B^2} = o(1)$

\item[(ii)] $\frac{K}{\log n} =O(1)$ and $(\log n)^{K+1} n^{-{K \underline{c}_A^3}/({\overline{c}_A\overline{c}_B})} =o(1)$.
\end{enumerate}
\end{assum}

Conditions (i)--(ii) in Assumption \ref{a-1} are the standard \emph{i.i.d.\ }and the quantile restrictions. Assumption \ref{a-1}(iii) requires some finite moment restrictions to achieve the probability bounds of various sample mean objects in the proof. Assumption \ref{a-2} allows conditional heteroskedasticity. Note that the eigenvalues of $A_{(m,k)}$ and $B_{(m,k)}$ are bounded and bounded away from zero for a given $(m,k)$. However, these bounds $(\underline{c}_{A(m,k)}, \underline{c}_{B(m,k)}, \overline{c}_{A(m,k)}, \overline{c}_{B(m,k)})$ can converge to zero or diverge to infinity as $n$ increases. The speed of convergence is restricted by Assumption \ref{a-2} (iv). These bounded eigenvalue restrictions are commonly imposed in the literature that studies the increasing dimension of parameters (see, e.g.\  \citet{portnoy1984asymptotic, portnoy1985asymptotic}).  Assumptions \ref{a-1}--\ref{a-2} are standard and similar to those in \citet{lu2015jackknife}. See the additional remarks therein.
Assumption \ref{a-3} imposes some regularity conditions on the number of \changed{potential regressors} $K_n$ and the sequence of the uniform bounds $(\underline{c}_A, \underline{c}_B, \overline{c}_A, \overline{c}_B)$. Different from the regularity condition of JMA in \citet{lu2015jackknife}, we need not restrict the growth rate of \changed{potential models} $M$ directly since $M_{(K_n,k)}$ is determined by $K_n$. However, $M_{(K_n,k)}$ increases very quickly at a factorial rate of $K_n$ and we need a stronger restriction on $K_n$. As noted in Assumption \ref{a-3}(ii), $K_n$ can increase at most the logarithmic rate of $n$. In the case of JMA, the number of regressors can increase at the polynomial rate if we set $\bar{k}=k_M=M$ in their notation. This is a trade-off in proving the uniform convergence results over a larger index set than that of JMA. We discuss this point in detail below in Theorem \ref{thm:uniform}. The second part of Assumption \ref{a-3}(ii) holds if \b{c}$%
_{A}^{3}/\bar{c}_{A}\bar{c}_{B}$ is bounded away from zero or converges to
zero at the slower rate than $\log (\log n)/\log n$ when $K$ increases at the rate of $\log n$.



First, we prove the convergence rate and the asymptotic normality of $\what{\Tht}_{(m,k)}$ when the dimension of parameter $k$ increases.

\begin{theorem}
\label{thm:pointwise} Suppose that Assumptions 1, 2, and 3(i) hold. Let $C_{(m,k)}$
denote an $l_{(m,k)} \times k$ matrix such that $C_{0}:=\lim_{n\rightarrow \infty
}C_{(m,k)}C_{(m,k)}^{\prime }$ exists and is positive definite, where $l_{(m,k)}
\in \left[ 1,k\right] $ is a fixed integer. Then,

\begin{enumerate}
\item[(i)] $\left\Vert \widehat{\Theta }_{(m,k)}-\Theta _{(m,k)}^{\ast
}\right\Vert =O_{p}\left( \sqrt{\frac{k}{n}}\right) $

\item[(ii)] $\sqrt{n}C_{(m,k)}V_{(m,k)}^{-1/2}\left[ \widehat{\Theta }%
_{(m,k)}-\Theta _{(m,k)}^{\ast }\right] \darrow N\left(
0,C_{0}\right) $\changed{.}
\end{enumerate}
\end{theorem}

This theorem provides an asymptotic theory for the quantile regression estimator when the model is misspecified and the number of parameters diverges to infinity as similarly seen in \citet{lu2015jackknife}. The convergence rate in (i) is a standard result when $k$ diverges as $n$ increases. To show the asymptotic normality with a diverging number of parameters, we also consider an arbitrary linear combination of $\widehat{\Theta}_{(m,k)}$ represented by $C_{(m,k)}$. The difference between two estimators, CSA and JMA, originates from the fact that CSA chooses the total number of the regressors $K_n$ first and the number of complete subset models $M_{(K_n,k)}$ follows automatically for each $k=1,\ldots,K_n$, whereas JSA selects the set of models $M_n$ (in their notation) in advance. Then, the size of regressors $k_m$ in case of JSA is determined by the sequence of models $m=1,\ldots, M_n$ chosen by a researcher. Although there are slight differences in the definition of $c_{A(m,k)}$ and $c_{B(m,k)}$ and their bounds from those in \citet{lu2015jackknife}, the proof of Theorem \ref{thm:pointwise} is identical to theirs, so is omitted.

We next turn our attention to the uniform convergence results of $\widehat{\Theta}_{i(m,k)}$ and $\widehat{\Tht}_{(m,k)}$. In addition to its own interest, the uniform convergence rates in the next theorem are required to prove to the asymptotic optimality of $\widehat{k}$.

\begin{theorem}
\label{thm:uniform} Suppose that Assumptions 1, 2\changed{,} and 3(ii) hold. Then,

\begin{enumerate}
\item[(i)] $\max_{1\leq i\leq n}\max_{1\leq k\leq K}\max_{1\leq m\leq
M}\left\Vert \widehat{\Theta }_{i(m,k)}-\Theta _{(m,k)}^{\ast }\right\Vert
=O_{p}\left( \sqrt{n^{-1}K\log n}\right) $

\item[(ii)] $\max_{1\leq k\leq K}\max_{1\leq m\leq M}\left\Vert \widehat{%
\Theta }_{(m,k)}-\Theta _{(m,k)}^{\ast }\right\Vert =O_{p}\left( \sqrt{%
n^{-1}K\log n}\right)$\changed{.}
\end{enumerate}
\end{theorem}

Since CSA is defined on the index sets of $m$ and $k$, the uniform convergence rates are defined over those sets, $m\in \{1,\ldots, M\}$ and $k\in \{1, \ldots, K\}$. In case of $\widehat{\Tht}_{i(m,k)}$, we need additional uniformity over $i\in\{1,\ldots,n\}$. As a result, the regularity conditions that control the growth rates of $K_n$ and $M_{(K_n,k)}$ are different from those of JMA in Assumption \ref{a-3} (ii). As discussed before, since the number of complete subsets increases at the factorial rate of $K_n$, we need a restriction on $K_n$ slightly stronger than that of JMA. We follow the proof strategy in \citet{lu2015jackknife} which extends the results of \citet{rice1984} by using the inequality in \citet{shibata1981optimal,shibata1982amendments}. To handle the different growth rates, we provide new technical lemmas. The proof of Theorem \ref{thm:uniform} as well as these lemmas are provided in the appendix.
Finally, the uniform convergence rates are expressed in terms of the sample size $n$ and the total number of regressors $K$ that goes to infinity as $n$ increases.

We next prove the prediction equivalence when we replace $M$ with $M_{max}$. Let $\mathcal{M}_{max}$ be a subset of $\{1,\ldots,M\}$ such that $M_{max}$ elements are randomly drawn. Define $\wtd{y}(k)$ to be the CSA quantile predictor using only $M_{max}$ models:
\eqs{
	\wtd{y}(k) := \frac{1}{M_{max}} \sum_{m' \in \mathcal{M}_{max}} x_{(m',k)}' \what{\Theta}_{(m',k)}.
}
Let $y^*_k := \lim_{M\rightarrow \infty} M^{-1} \sum_{m=1}^M E\lt[x_{(m,k)}' \Tht^*_{(m,k)}\rt] < \infty$. We show the validity of $M_{max}$ in the following theorem:
\begin{theorem}\label{thm:subsample}
Suppose that Assumptions 1--3 hold. Let $M_{max} \rightarrow \infty$ and $K/M_{max} \rightarrow 0$ as $n \rightarrow \infty$. Furthermore, we assume that 
\eqs{
& \max_{1\le k \le K} M^{-1} \sum_{m=1}^M x_{(m,k)}' \Theta^{*}_{(m,k)} - y^*_k =o_p(1) \\
& \max_{1\le k \le K}\max_{1 \le  m \le M} \Vert x_{(m,k)} \Vert =O_p(1).
}
Then, we have
\eqs{
	\max_{1 \le k \le K}\lt\vert \what{y}(k) - \wtd{y}(k)\rt\vert =o_p(1).
}
\end{theorem}

The rate requirement for $M_{max}$ is mild and $M_{max}=O(n^{1/2})$ would work given $K=O(\log n)$. The uniform boundedness assumption on $\Vert x_{(m,k)} \Vert$ is weak and holds easily in most applications.
We have some remarks on the uniform convergence assumption of the model average with the pseudo-true parameter $\Theta^*_{(m,k)}$. 
Let $z_{(m,k)} = x_{(m,k)}'\Tht^*_{(m,k)} - E\lt[x_{(m,k)}' \Tht^*_{(m,k)}\rt]$. 
Note that $k$ is discrete and the functional class size over $k$ is small.
Thus, it depends on the dependent structure of $z_{(m,k)}$ to hold the uniform law of large numbers. 
For example, consider the following maximal inequality: for $\dt>0$,
\eqs{
	 P\lt( \max_{1 \le k \le K} \lt\vert M^{-1} \sum_{m=1}^M z_{(m,k)} \rt\vert > \dt \rt) 
	& \le K \max_{1 \le k \le K}P\lt( \lt\vert M^{-1} \sum_{m=1}^M z_{(m,k)}  \rt\vert > \dt \rt) \\
	& \le \frac{K}{M} \max_{1 \le k \le K} \frac{E[\sum_{m=1}^M z_{m,k} ]^2}{M \dt^2 },
}
where the second line holds from the Markov inequality. 
Since $K/M =o(1)$, a sufficient condition for the uniform convergence is $\max_{1 \le k \le K}E[\sum_{m=1}^M z_{m,k}]^2/M =O(1)$. 
If $z_{(m,k)}$ is covariance stationary over $m$ for all $k$, then the sufficient \changed{condition} becomes the absolute summability condition $\max_{1\le k \le K} \sum_{j=0}^{\infty} \lt\vert E[z_{(m,k)} z_{(m+j,k)}] \rt\vert < \infty$. See, e.g. \citet{fazekas2001general} for more general conditions on \changed{the} partial sums in a different dependent structure.

We now prove the asymptotic optimality of $\widehat{k}$ in the sense of \citet{li1987asymptotic}. Following \citet{lu2015jackknife}, we use the final prediction error (FPE, or the out-of-sample quantile prediction error) as a criterion to evaluate the prediction performance:
\begin{align*}
FPE_n(k) := E\left[\rho_{\tau} \left(y- \frac{1}{M} \sum_{m=1}^{M}
X_{(m,k)}^{\prime} \widehat {\Theta}_{(m,k)} \right)\vert \mathcal{D}_n %
\right],
\end{align*}
where $\mathcal{D}_n:=\{(y_i,x_i):i=1,\ldots,n\}$ is a sample. The next theorem shows that $\what{k}$ is asymptotically equivalent to the infeasible best subset size choice that is defined as a minimizer of $FPE(k)$.
\begin{theorem}\label{thm:opt}
Suppose that Assumptions 1--3. Then,
\begin{equation*}
\frac{FPE(\widehat{k})}{\inf_{k\in \mathcal{K}}FPE(k)}\overset{p}{\rightarrow }1\changed{,}
\end{equation*}
where $\mathcal{K}:=\left\{ 1,...,K_{n}\right\} .$
\end{theorem}
A similar optimality concept has been adopted in the context of the weighted average estimator (e.g.\ \citet{hansen2007least}, \citet{hansen2012jackknife}, and \citet{lu2015jackknife}) and in the context of the IV estimator (e.g.\ \citet{donald2001choosing}, \citet{kuersteiner2010constructing}, and \citet{lee2018complete}). 
Different from JMA, CSA considers the complete subsets given $(K_n,k)$ and does not require the pre-selection of models to be considered nor the order of models. Thus, the optimality result is also independent of the initial model selection/ordering issue once the total number of regressors is given. 
The index set $\mathcal{K}$ of CSA is discrete while that of JMA or the jackknife model averaging in \citet{hansen2012jackknife} is compact. All require the finite moment condition similar to Assumption (A.1) in \citet{li1987asymptotic} which is assured by Assumption 1 (iii) above. 
The idea of complete subset averaging has been adopted in the forecasting literature (e.g.\ \citet*{elliott2013complete, elliott2015complete}, \citet*{rapach2010out})\changed{. This }is the first formal result to show the optimality of the subset size selection.

Finally, we compare the performance of the nonstochastic equal weight with that of the optimal weight.
In the mean prediction context, it has been observed that a simple arithmetic mean, i.e.~the equal weight, outperforms the \emph{estimated} optimal weight.
This empirical phenomenon is known as \changed{the} `forecast combination puzzle' and some formal explanations under the mean squared error are provided by \citet{smith2009simple}, \citet{elliott2011averaging}, and \citet*{claeskens2016forecast}, to name a few.
Heuristically speaking, it happens when the estimation error of the optimal weight is large enough to dominate the efficiency loss caused by the equal weight.
We extend this result to the class of smooth expected loss functions. This is crucial in our analysis since the check function $\rho_{\tau}(\cdot)$ does not give a closed-form solution, which is different from the mean squared error used in the existing literature.

We consider the following simplified framework to focus on the main idea.
Let be $\hat{y}_{1}, \ldots, \hat{y}_{M}$ be predictors for $y$ based on $M$ different models.
For example, we can think of $\hat{y}_{m} = X'_{(m,k)}\what{\Theta}_{(m,k)}$ for any given $k$.
Let $w$ be \changed{an} $M$-dimensional weight vector combining the $M$ predictors. We consider only positive weights with $1_M'w=1$, where $1_M$ is \changed{an} $M$-dimensional unit vector. Let $\hat{y}:=(\hat{y}_1,\ldots, \hat{y}_M)'$ and $e_m:=y - \hat{y}_m$ be the prediction error of $\hat{y}_m$ and $e:=(e_1,\ldots,e_M)'$ be a vector of \changed{these prediction errors}. We define the prediction error of the combined predictor as $e_c(w):=y-w'\hat{y}=w'(1_M \cdot y - \hat{y})=w'e$\changed{.} Then, we can define an optimal weight $w^*$ as
\eqs{
	w^* = \argmin_{w \in \Delta^{M-1}} F(w),
}where $\Delta^{M-1}$ is the standard $(M-1)$-simplex and $F(w):=E[L(w;e_c)]$ is an expected loss function. For example, the mean squared error in \citet{elliott2011averaging} can be written in terms of the quadratic loss function: $F(w)=E[e_c^2]=E[w'e e'w]=w'\Sigma w$, where $\Sigma=E[e e']$. The quantile prediction error adopted in this paper can be written in terms of the check function: $F(w) = E[\rho_{\tau}(e_c)] = E[\rho_{\tau}(y-w'\hat{y})]$. Let $\bar{w}:=M^{-1}1_M$ be an equal-weight vector and $\hat{w}$ be an estimator for $w^*$ with $\hat{\eta} := \hat{w} - w^*$. To illustrate our main point\changed{,} we further impose that $E[\hat{\eta}]=0$ and $\max_m Var(\hat{\eta}_m) = \bar{\sigma}_{\eta}^2 > 0$.

\begin{theorem}\label{thm:prediction-efficiency-bound}
	Suppose that $F(w)$ is twice differentiable on $\Delta^{M-1}$ and that $\sup_{w \in \Delta^{M-1}} \left\Vert \triangledown_2 F(w) \right\Vert \le C <\infty$ uniformly in $M$. Let $\bar{\ld}_{max} :=\lim\sup_{M} \sup_w \ld_{max}\lt(\triangledown_2 F(w)\rt)$.
	\ben
		\item[(i)] $\left\vert F(\bar{w}) - F(w^*) \right\vert  \le 2^{-1}\bar{\lambda}_{max} \left(1 + {3}{M}^{-1}\right)$
        \item[(ii)] \changed{ E$\left\vert F(\hat{w}) - F(w^*) \right\vert  \le 2^{-1}\bar{\lambda}_{max} M \bar{\sigma}_{\eta}^2$.}
	\een
\end{theorem}
We have some remarks.
First, it shows that the equal weight $\bar{w}$ may work better than the estimated optimal weight $\hat{w}$ when we average many models, i.e.\ when $M$ is large.
Compared to the optimal prediction error $F(w^*)$, the efficiency loss by $\bar{w}$ is bounded by $2^{-1}\bar{\ld}_{max}(1+3M^{-1})$, which converges to $2^{-1}\bar{\ld}_{max}$ for large enough $M$.
On the contrary, the upper bound of the \changed{mean} efficiency loss by $\hat{w}$ diverges as $M$ increases.
We admit that these upper bounds only reflect the worst case scenario.
However, it confirms the intuition formally that the equal weight can outperform the \emph{estimated} optimal weight under the class of smooth expected loss functions.
Second,
the prediction error of $\bar{w}$ under a quadratic loss function converges to the optimal prediction error as $M$ increases.
The same result is also proved in Proposition 1 in \citet{elliott2011averaging}.
Different from his result, it does not require \changed{decomposing} the prediction error into the common component and the idiosyncratic component.
This result is summarized in Corollary \ref{cor:equivalence} below.
Third, to achieve the optimality, the \changed{estimation errors of the weight $\hat{w}$, $\{\hat{\eta}_m\}$},  should vanish fast enough. Let $\bar{\sigma}_{\eta}^2=O(c_n)$. A sufficient condition for the optimality is $c_n=o(M_n^{-1})$. For example, if $c_n$ is a parametric rate, $n^{-1/2}$, then $M_n$ should be bounded by $o(n^{1/2})$.  When $M_n=O(\log n)$, for example, this condition is satisfied.
However, if \changed{$M_n$ increases too fast,} \changed{then $M_n\bar{\sigma}_{\eta}^2$ will diverge} and $\hat{w}$ may work worse than $\bar{w}$.\footnote{We thank an anonymous referee and Co-editor for pointing out this intuition. Also, note that it is one \emph{sufficient} condition. It is still possible that there exists a different set of conditions that guarantee the optimality.}
Fourth,
$\lt\Vert \triangledown_2 F(w) \rt\Vert = (\sum_{m=1}^M \ld_m^2 )^{1/2}$, where $\{\ld_m\}$ are eigenvalues of $\triangledown_2 F(w)$ since $\triangledown_2 F(w)$ is symmetric.
Thus, the uniform bound $C$ exists if $\{\ld_m\}$ is absolutely summable, $\sum_{m=1}^{\infty} \lt\vert \ld_m \rt\vert < \infty$.
Fifth,
if we restrict our attention to the expected check function adopted in this paper,
$F(w)$ is twice differentiable if the conditional density $f(y|\hat{y})$ is smooth for all $\hat{y}$. From Theorem 1 in \citet*{angrist2006quantile}, we have
\eq{
	F(w) = E\lt[ \bar{\omega}_{\tau}(\hat{y},w)(w'\hat{y} - Q_{\tau}(y|\hat{y}))^2 \rt]\changed{,} \label{eq:obj-mean-expression}
}where $Q_{\tau}(y|\hat{y})$ is the conditional quantile function of $y$ given $\hat{y}$ and $\bar{\omega}_{\tau}(\hat{y},w):=\int_0^1 (1-u)\cdot f(u\cdot w'\hat{y} + (1-u) \cdot Q_{\tau}(y|\hat{y})|\hat{y})du$. Thus, the smoothness of $F(w)$ is implied by the twice differentiability of $f(y|\hat{y})$.
Finally,
equation \eqref{eq:obj-mean-expression} shows that CSA would not work well if we include many irrelevant models. Similar to the quantile regression specification error in \citet{angrist2006quantile}, we call $\lt(w'\hat{y} - Q_{\tau}(y|\hat{y})\rt)$ the quantile prediction specification error. If there are many irrelevant models, the optimal weight $\omega^*$ would be sparse, i.e.\ many elements of $\omega^*$ would be zeros. In such a case, CSA with $\bar{\omega}=M^{-1}1_M$ results in a larger quantile prediction specification error given $M$ and $n$. For example, if there is only one relevant regressor and all other coefficients $\theta_j(\tau)$ \changed{equal} zero besides one, the complete subsets will be composed of many irrelevant models. As we will see in the simulations studies in the next section, CSA does not perform well under this situation. Therefore, a pre-screening process is desirable to achieve a satisfactory result of CSA.
\begin{corollary} \label{cor:equivalence}
	Suppose that we have a quadratic loss function, $L(w;e_c)=(e_c(w))^2$ and that $\ld_{max}(\Sigma) < \infty$ uniformly in $M$, where $\Sigma:=E[ee']$. Then, we have
	\eqs{
		\lt\vert F(\bar{w}) - F(w^*) \rt\vert \rightarrow 0 \mbox{ as } M \rightarrow \infty.
	}
\end{corollary}

\section{Monte Carlo Simulations}\label{sc:simulations}
In this section, we investigate the finite sample performance of the proposed
estimator in simple Monte Carlo experiments. We consider two categories of the simulation designs: (i) all candidate models are misspecified, and (ii) candidate models include the true model. 

First, we adopt the following data generating process (DGP):
\begin{align}
y_i = \theta \sum_{j=1}^{1000} j^{-1} x_{ij} + \varepsilon _i, \label{eq:sim design}
\end{align}%
where $x_{i1}=1$ and $(x_{i2},\ldots, x_{i1000})$ follows a multivariate normal distribution, $N(0,\Sigma)$ with $\Sigma_{jk}=\rho_x$ if $j\neq k$ and 1 if $j=k$.
Therefore, the regressors are possibly dependent \changed{on} each other, which is a more general feature of the design than the existing literature, see, e.g., \citet{hansen2007least} and \citet{lu2015jackknife}. 
The term $\varepsilon _i$ follows $N(0,1)$ independent of $x_{ij}$. The sample is \emph{i.i.d. }over $i$.
The population $R^2:=(Var(y_i)-Var(\varepsilon _i))/Var(y_i)$ is controlled by $\theta$.
We consider two sample sizes, $n=50, 150$.
The number of potential regressors is set to $K=4\log(n)$, which is 15 and 20, respectively.
Note that all candidate models are misspecified since there remain many missing regressors in the sample.
We consider various DGPs by combining different $R^2 = \{0.1, \ldots, 0.9\}$, $\tau = \{0.1, \ldots, 0.9\}$, and $\rho_x = \{0.0, 0.1, 0.2, \ldots, 0.9\}$. We consider 38 different DGPs in total and estimate 74 different quantile models.

We compare the performance of the proposed Complete Subset Averaging estimator
(CSA) with the Jackknife Model Averaging estimator \changed{(JMA)} in \citet{lu2015jackknife}, the $\ell_1$-penalized quantile regression (L1QR) in \citet*{belloni2011L1}, the bootstrap aggregating methods (BAG) in \citet{breiman1996bagging} and $\ell_2$-penalized quantile regression. L1QR and L2QR are also called the lasso and the ridge regression in the mean regression setup. The set of models used for JMA is constructed in an encompassing way, e.g.\ $\{x_{i1}\}, \{x_{i1},x_{i2}\},\ldots, \{x_{i1},\ldots, x_{i20}\}$. For CSA, we set the maximum submodels to $M_{max}=100$. Thus, we draw 100 models randomly from the complete subsets of size $k$ if $M=K!/(k!(K-k)!)$ is bigger than 100. Furthermore, we reduce some computational burden by applying 10-fold cross-validation when $n=150$. The tuning parameter of L1QR is chosen by Equation (2.7) in \citet{belloni2011L1}. The bootstrap size of BAG is set to be 1000. The tuning parameter of L2QR is chosen by 10-fold cross-validation over the set $\{0.01, 0.05, 0.1, 0.5, 1.0\}$ which is constructed after some pre-simulation studies. 

To compare the performance, we first compute $\mbox{FPE}(r)$ for each replication $r=1,\ldots,R$ as follows. After estimating the model with $n$ in-sample observations, we \changed{generate} additional 100 out-of-sample observations. Then, $\mbox{FPE}(r)$ is calculated by
\begin{align*}
\mbox{FPE}(r) := \frac{1}{100} \sum_{s=1}^{100} \rho_{\tau} \left(y_s - \hat{y}_s\right)\changed{,}
\end{align*}%
where $\hat{y}_x$ is a predicted value by each method.
Then, we construct the following three comparison measures:
\eqs{
  \mbox{Average FPE}_A     &:= R^{-1}\sum_{r=1}^R \mbox{FPE}(r)_A \\
  \mbox{Winning Ratio}_{A}   &:= R^{-1}\sum_{r=1}^R 1\{\mbox{FPE}(r)_A < \mbox{FPE}(r)_B, \ldots, \mbox{FPE}(r)_A < \mbox{FPE}(r)_E \} \\
  \mbox{Loss to CSA}_{A} &:= R^{-1}\sum_{r=1}^R 1\{\mbox{FPE}(r)_{CSA} < \mbox{FPE}(r)_A\},
}where each subscript denotes generic notation for a forecasting method.
Note that the loss to CSA ratio provides more direct binary comparison of each method to CSA. 
We set the total number of replications $R=1000$.  
\begin{figure}[tbp]
\caption{Prediction Errors over $R^2$}
\label{fig:R2}\centering
\vskip10pt
\resizebox{14cm}{!}{
\begin{tabular}{cc}
\multicolumn{2}{c}{$\tau=0.5$} \\
$n=50$ & $n=150$ \\
\includegraphics[scale=0.6]{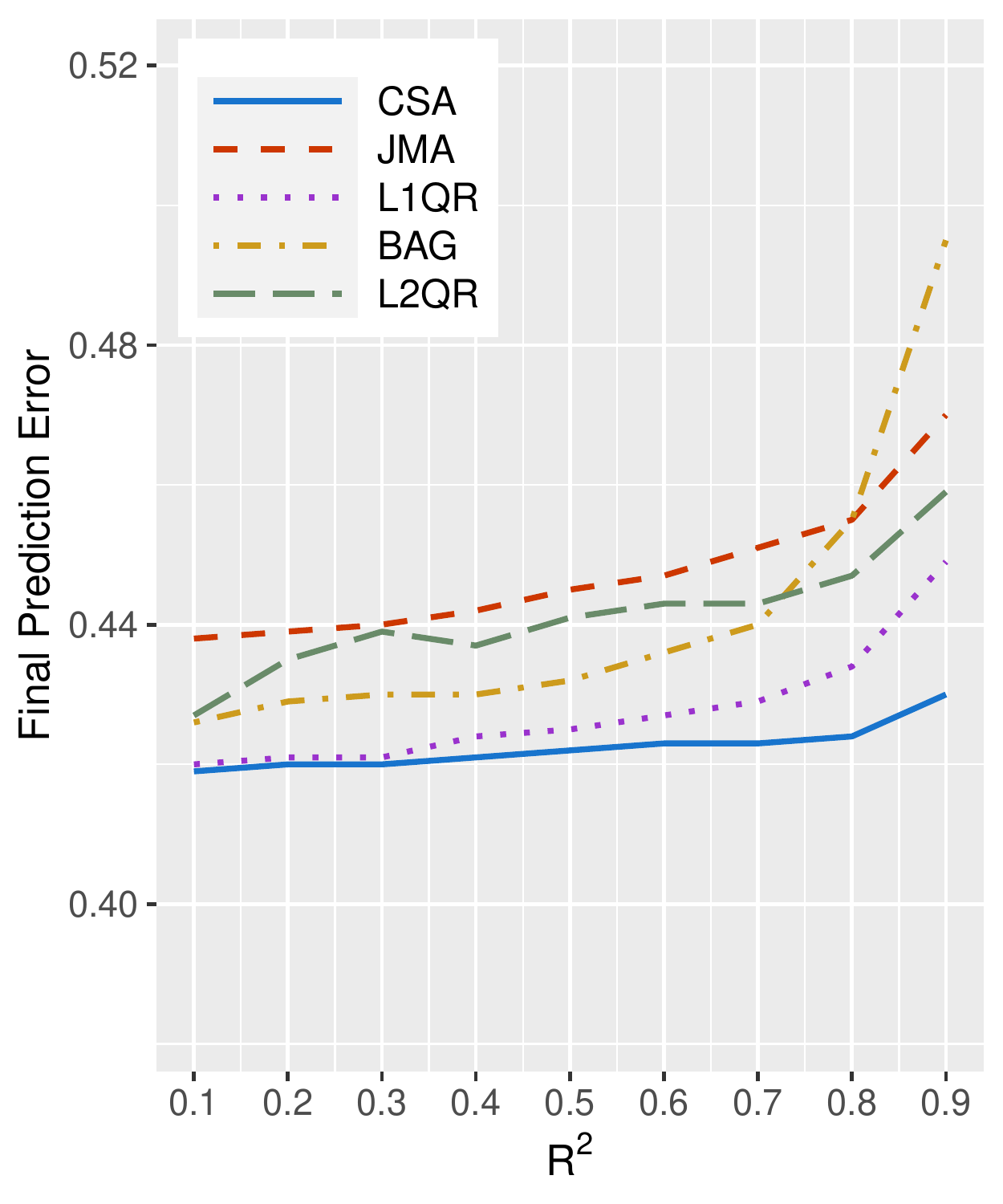} & %
\includegraphics[scale=0.6]{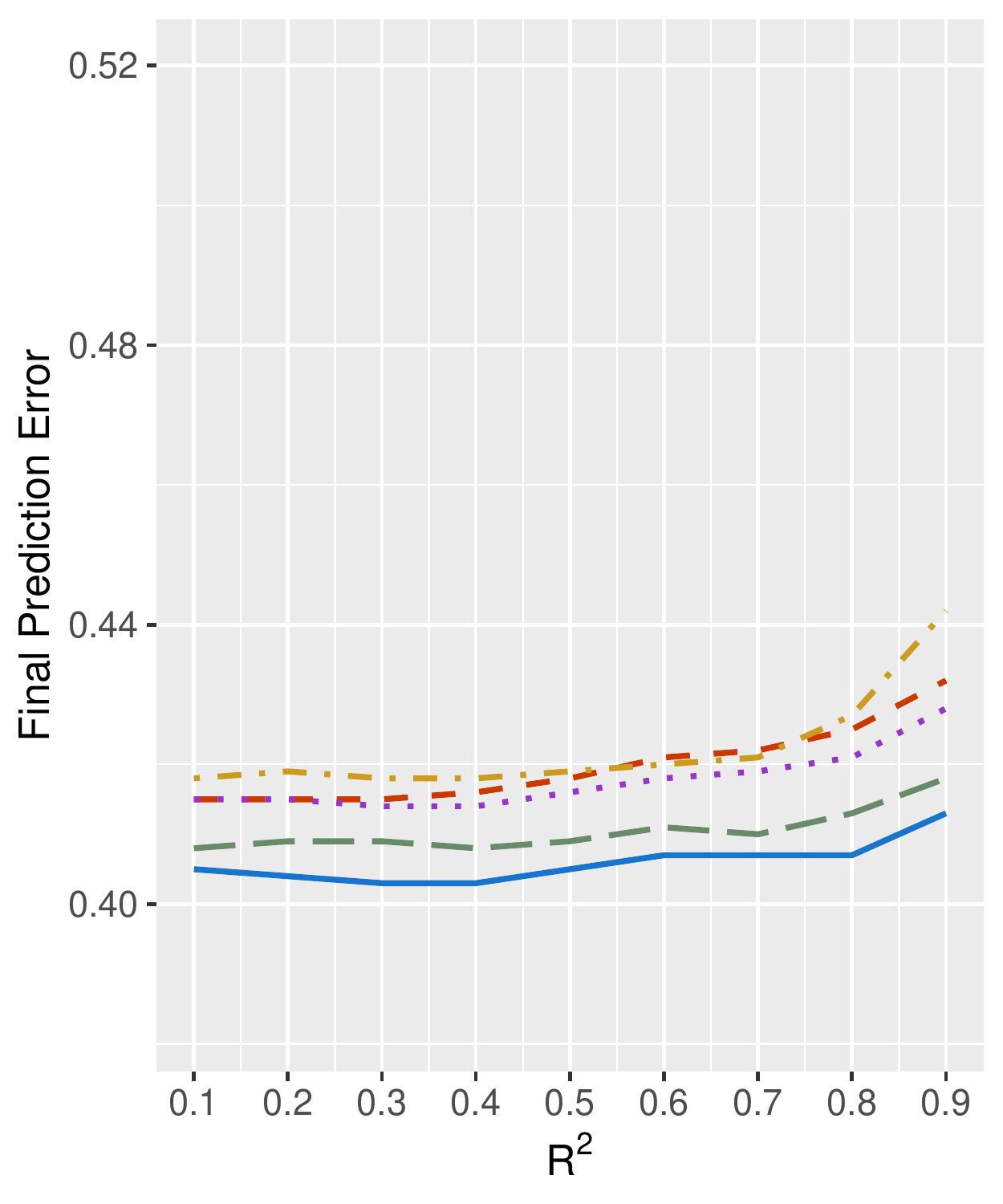} \\
&  \\
\multicolumn{2}{c}{$\tau=0.1$} \\
$n=50$ & $n=150$ \\
\includegraphics[scale=0.6]{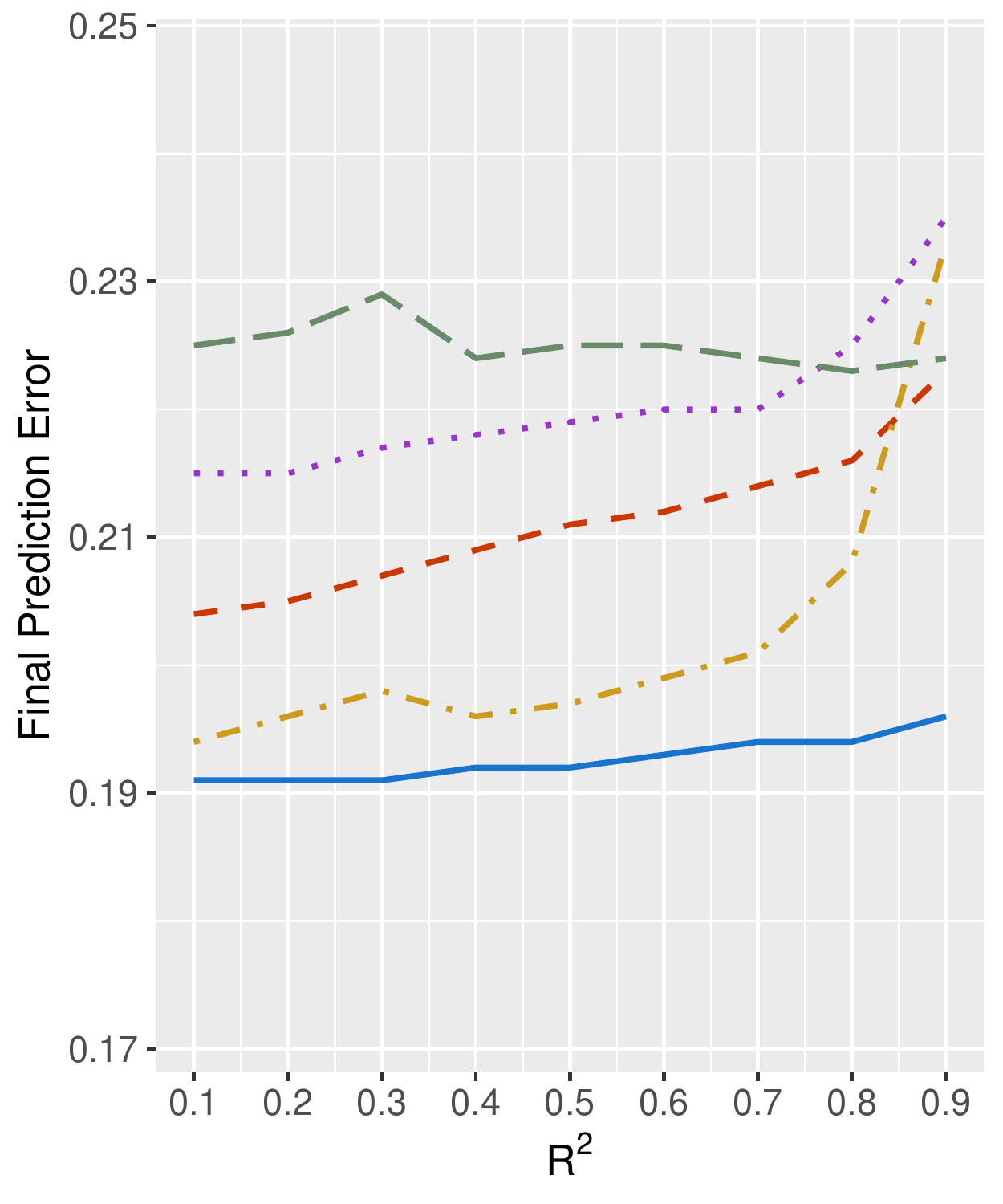} & %
\includegraphics[scale=0.6]{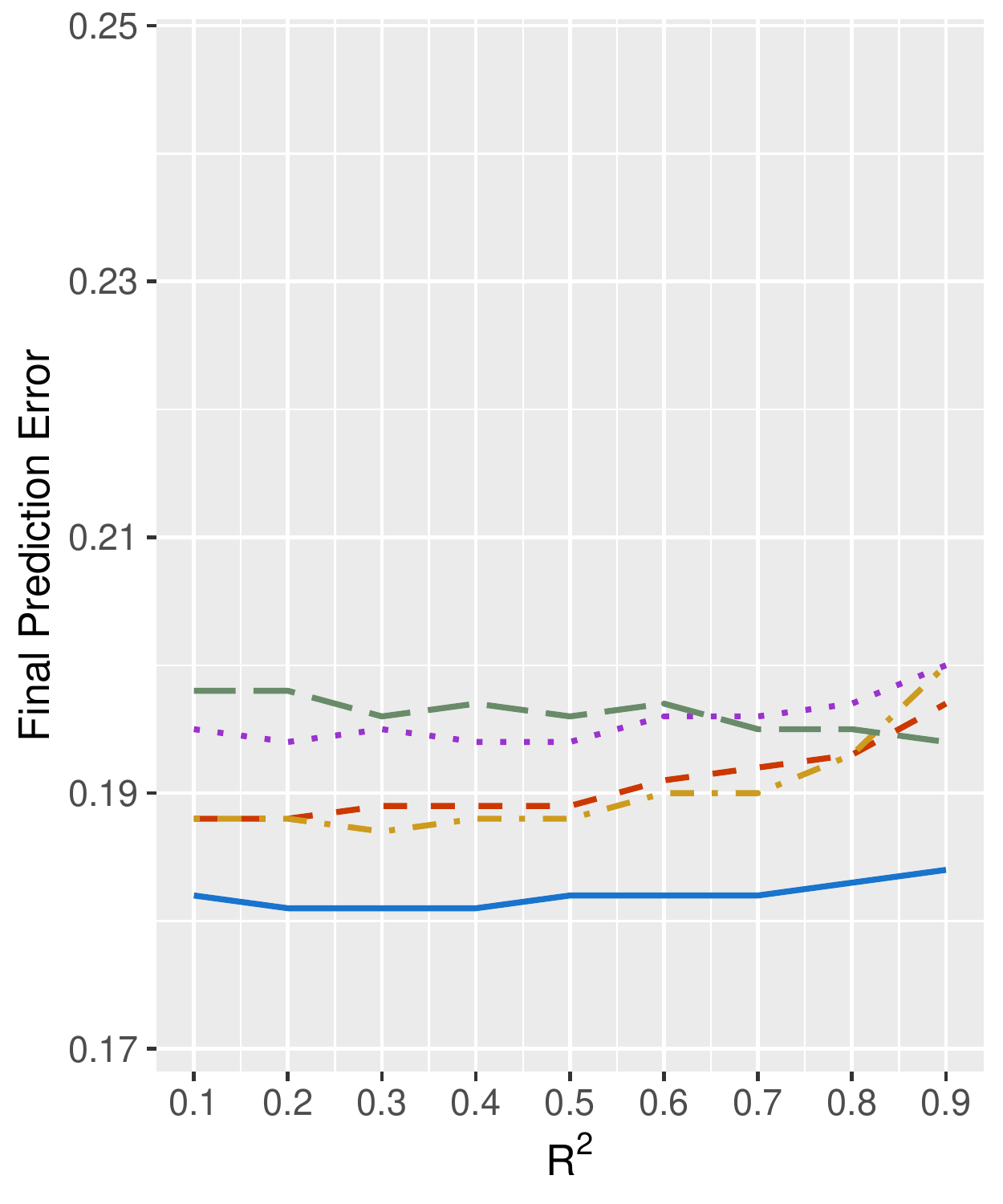}%
\end{tabular}
}
\end{figure}

\begin{table}[tbp]
\centering
\begin{threeparttable}
\caption{Simulation Results over Various $R^2$: $\tau=0.5$}\label{tb:R2-0.5quant}
\begin{tabular}{lccccccccccc}
\\
\hline
      &  \multicolumn{5}{c}{$n=50$}    &  & \multicolumn{5}{c}{$n=150$}   \\
\cline{2-6} \cline{8-12}
$R^2$ &  CSA  & JMA & L1QR & BAG & L2QR & & CSA  & JMA & L1QR & BAG & L2QR \\
\hline
\multicolumn{12}{c}{\underline{Average FPE}}\\
  \multirow{2}{*}{0.1} 	&  0.419 & 0.438 & 0.420 & 0.426 & 0.427 &  & 0.405 & 0.415 & 0.415 & 0.418 & 0.408 \\ 
      					&  (0.043) & (0.047) & (0.037) & (0.035) & (0.048) &  & (0.033) & (0.035) & (0.033) & (0.033) & (0.033) \\ 
  \multirow{2}{*}{0.2}	&  0.420 & 0.439 & 0.421 & 0.429 & 0.435 &  & 0.404 & 0.415 & 0.415 & 0.419 & 0.409 \\ 
      					&  (0.042) & (0.046) & (0.036) & (0.035) & (0.051) &  & (0.032) & (0.034) & (0.033) & (0.033) & (0.033) \\ 
  \multirow{2}{*}{0.3}	&  0.420 & 0.440 & 0.421 & 0.430 & 0.439 &  & 0.403 & 0.415 & 0.414 & 0.418 & 0.409 \\ 
   	  					&  (0.041) & (0.045) & (0.036) & (0.035) & (0.053) &  & (0.032) & (0.033) & (0.033) & (0.032) & (0.032) \\ 
  \multirow{2}{*}{0.4}	&  0.421 & 0.442 & 0.424 & 0.430 & 0.437 &  & 0.403 & 0.416 & 0.414 & 0.418 & 0.408 \\ 
      					&  (0.042) & (0.047) & (0.035) & (0.035) & (0.049) &  & (0.032) & (0.034) & (0.033) & (0.032) & (0.031) \\ 
  \multirow{2}{*}{0.5}	&  0.422 & 0.445 & 0.425 & 0.432 & 0.441 &  & 0.405 & 0.418 & 0.416 & 0.419 & 0.409 \\ 
      					&  (0.042) & (0.046) & (0.035) & (0.035) & (0.053) &  & (0.032) & (0.034) & (0.033) & (0.032) & (0.032) \\ 
  \multirow{2}{*}{0.6}	&  0.423 & 0.447 & 0.427 & 0.436 & 0.443 &  & 0.407 & 0.421 & 0.418 & 0.420 & 0.411 \\ 
      					&  (0.041) & (0.045) & (0.035) & (0.036) & (0.055) &  & (0.032) & (0.034) & (0.033) & (0.032) & (0.033) \\ 
  \multirow{2}{*}{0.7}	&  0.423 & 0.451 & 0.429 & 0.440 & 0.443 &  & 0.407 & 0.422 & 0.419 & 0.421 & 0.410 \\ 
      					&  (0.043) & (0.046) & (0.036) & (0.036) & (0.052) &  & (0.031) & (0.033) & (0.033) & (0.032) & (0.032) \\ 
  \multirow{2}{*}{0.8}	&  0.424 & 0.455 & 0.434 & 0.455 & 0.447 &  & 0.407 & 0.425 & 0.421 & 0.427 & 0.413 \\ 
      					&  (0.042) & (0.046) & (0.036) & (0.038) & (0.051) &  & (0.031) & (0.033) & (0.033) & (0.032) & (0.032) \\ 
  \multirow{2}{*}{0.9}	&  0.430 & 0.470 & 0.449 & 0.495 & 0.459 &  & 0.413 & 0.432 & 0.428 & 0.442 & 0.418 \\ 
      					&  (0.043) & (0.048) & (0.038) & (0.049) & (0.055) &  & (0.033) & (0.035) & (0.034) & (0.035) & (0.033) \\ 
\multicolumn{12}{c}{\underline{Winning Ratio}}\\
  0.1 & 27.8\% & 8.6\% & 14.9\% & 19.5\% & 29.2\% &  & 34.2\% & 10.1\% & 6.8\% & 11.3\% & 37.6\% \\ 
  0.2 & 29.6\% & 7.7\% & 17.2\% & 19.7\% & 25.8\% &  & 35.9\% & 10.1\% & 7.7\% & 11.8\% & 34.5\% \\ 
  0.3 & 33.6\% & 6.9\% & 16.2\% & 20.9\% & 22.4\% &  & 39.4\% & 8.9\% & 7.9\% & 10.9\% & 32.9\% \\ 
  0.4 & 34.3\% & 6.4\% & 14.4\% & 21.0\% & 23.9\% &  & 38.5\% & 7.4\% & 7.7\% & 12.5\% & 33.9\% \\ 
  0.5 & 35.8\% & 5.9\% & 16.3\% & 18.0\% & 24.0\% &  & 38.4\% & 8.8\% & 7.2\% & 13.4\% & 32.2\% \\ 
  0.6 & 36.6\% & 7.1\% & 15.6\% & 17.7\% & 23.0\% &  & 40.1\% & 7.4\% & 7.5\% & 13.0\% & 32.0\% \\ 
  0.7 & 38.8\% & 4.4\% & 15.5\% & 16.0\% & 25.3\% &  & 40.4\% & 6.3\% & 7.1\% & 12.3\% & 33.9\% \\ 
  0.8 & 47.2\% & 4.4\% & 12.7\% & 11.1\% & 24.6\% &  & 43.2\% & 6.7\% & 6.2\% & 10.0\% & 33.9\% \\ 
  0.9 & 54.1\% & 4.2\% & 9.9\% & 3.6\% & 28.2\% &  & 43.0\% & 5.3\% & 6.1\% & 6.7\% & 38.9\% \\ 
\multicolumn{12}{c}{\underline{Loss to CSA}}\\
  0.1 & NA & 77.3\% & 59.6\% & 57.0\% & 55.6\% &  & NA & 77.8\% & 81.8\% & 65.4\% & 52.5\% \\ 
  0.2 & NA & 77.8\% & 59.4\% & 58.4\% & 59.8\% &  & NA & 79.1\% & 82.6\% & 64.2\% & 55.2\% \\ 
  0.3 & NA & 79.8\% & 62.2\% & 61.1\% & 61.5\% &  & NA & 80.4\% & 82.8\% & 66.3\% & 57.3\% \\ 
  0.4 & NA & 79.3\% & 65.1\% & 60.0\% & 62.3\% &  & NA & 82.8\% & 83.2\% & 65.2\% & 54.2\% \\ 
  0.5 & NA & 80.7\% & 65.6\% & 61.6\% & 62.1\% &  & NA & 81.9\% & 81.8\% & 65.5\% & 55.5\% \\ 
  0.6 & NA & 82.4\% & 67.2\% & 64.3\% & 64.7\% &  & NA & 82.7\% & 82.4\% & 66.1\% & 57.1\% \\ 
  0.7 & NA & 84.7\% & 69.8\% & 64.5\% & 62.3\% &  & NA & 85.8\% & 82.9\% & 65.2\% & 54.8\% \\ 
  0.8 & NA & 86.4\% & 75.5\% & 73.2\% & 65.4\% &  & NA & 85.6\% & 85.4\% & 69.5\% & 56.5\% \\ 
  0.9 & NA & 89.2\% & 82.6\% & 86.0\% & 68.0\% &  & NA & 87.2\% & 84.6\% & 74.5\% & 54.3\% \\ 
\hline
\end{tabular}
    \begin{tablenotes}
      \small
      \item Notes: The standard error of the average FPE is denoted inside the parentheses.
    \end{tablenotes}
\end{threeparttable}
\end{table}

\begin{table}[tbp]
\centering
\caption{Simulation Results over Various $R^2$: $\tau=0.1$}\label{tb:R2-0.1quant}
\begin{tabular}{lccccccccccc}
\\
\hline
      &  \multicolumn{5}{c}{$n=50$}    &  & \multicolumn{5}{c}{$n=150$}   \\
\cline{2-6} \cline{8-12}
$R^2$ &  CSA  & JMA & L1QR & BAG & L2QR & & CSA  & JMA & L1QR & BAG & L2QR \\
\hline
\multicolumn{12}{c}{\underline{Average FPE}}\\
  \multirow{2}{*}{0.1} 	&  0.191 & 0.204 & 0.215 & 0.194 & 0.225 &  & 0.182 & 0.188 & 0.195 & 0.188 & 0.198 \\ 
      					&  (0.028) & (0.036) & (0.040) & (0.023) & (0.046) &  & (0.019) & (0.021) & (0.025) & (0.020) & (0.026) \\ 
  \multirow{2}{*}{0.2}	&  0.191 & 0.205 & 0.215 & 0.196 & 0.226 &  & 0.181 & 0.188 & 0.194 & 0.188 & 0.198 \\ 
      					&  (0.027) & (0.035) & (0.039) & (0.024) & (0.050) &  & (0.019) & (0.021) & (0.024) & (0.021) & (0.027) \\ 
  \multirow{2}{*}{0.3}	&  0.191 & 0.207 & 0.217 & 0.198 & 0.229 &  & 0.181 & 0.189 & 0.195 & 0.187 & 0.196 \\ 
   	  					&  (0.029) & (0.036) & (0.040) & (0.025) & (0.049) &  & (0.019) & (0.021) & (0.025) & (0.020) & (0.025) \\ 
  \multirow{2}{*}{0.4}	&  0.192 & 0.209 & 0.218 & 0.196 & 0.224 &  & 0.181 & 0.189 & 0.194 & 0.188 & 0.197 \\ 
      					&  (0.028) & (0.037) & (0.039) & (0.023) & (0.047) &  & (0.019) & (0.021) & (0.024) & (0.021) & (0.027) \\ 
  \multirow{2}{*}{0.5}	&  0.192 & 0.211 & 0.219 & 0.197 & 0.225 &  & 0.182 & 0.189 & 0.194 & 0.188 & 0.196 \\ 
      					&  (0.029) & (0.036) & (0.040) & (0.023) & (0.046) &  & (0.019) & (0.021) & (0.025) & (0.020) & (0.026) \\ 
  \multirow{2}{*}{0.6}	&  0.193 & 0.212 & 0.220 & 0.199 & 0.225 &  & 0.182 & 0.191 & 0.196 & 0.190 & 0.197 \\ 
      					&  (0.029) & (0.037) & (0.042) & (0.024) & (0.047) &  & (0.018) & (0.021) & (0.024) & (0.020) & (0.026) \\ 
  \multirow{2}{*}{0.7}	&  0.194 & 0.214 & 0.220 & 0.201 & 0.224 &  & 0.182 & 0.192 & 0.196 & 0.190 & 0.195 \\ 
      					&  (0.032) & (0.038) & (0.040) & (0.025) & (0.046) &  & (0.018) & (0.021) & (0.024) & (0.021) & (0.025) \\ 
  \multirow{2}{*}{0.8}	&  0.194 & 0.216 & 0.225 & 0.208 & 0.223 &  & 0.183 & 0.193 & 0.197 & 0.193 & 0.195 \\ 
      					&  (0.028) & (0.035) & (0.041) & (0.027) & (0.046) &  & (0.018) & (0.022) & (0.025) & (0.020) & (0.024) \\ 
  \multirow{2}{*}{0.9}	&  0.196 & 0.223 & 0.235 & 0.233 & 0.224 &  & 0.184 & 0.197 & 0.200 & 0.200 & 0.194 \\ 
      					&  (0.027) & (0.035) & (0.041) & (0.032) & (0.046) &  & (0.018) & (0.022) & (0.025) & (0.021) & (0.023) \\ 
\multicolumn{12}{c}{\underline{Winning Ratio}}\\
  0.1 & 38.8\% & 10.0\% & 8.0\% & 34.2\% & 9.0\% &  & 38.4\% & 12.8\% & 8.3\% & 26.7\% & 13.8\% \\ 
  0.2 & 41.2\% & 12.5\% & 8.5\% & 31.2\% & 6.6\% &  & 37.0\% & 12.8\% & 8.8\% & 28.0\% & 13.4\% \\ 
  0.3 & 43.0\% & 9.5\% & 8.1\% & 32.5\% & 6.9\% &  & 37.5\% & 11.1\% & 8.8\% & 27.4\% & 15.2\% \\ 
  0.4 & 42.0\% & 7.9\% & 7.8\% & 32.3\% & 10.0\% &  & 41.2\% & 11.2\% & 8.2\% & 26.0\% & 13.4\% \\ 
  0.5 & 43.7\% & 7.7\% & 8.9\% & 31.5\% & 8.2\% &  & 38.6\% & 10.7\% & 9.2\% & 26.7\% & 14.8\% \\ 
  0.6 & 45.1\% & 7.8\% & 7.9\% & 27.7\% & 11.5\% &  & 40.5\% & 11.3\% & 8.7\% & 23.9\% & 15.6\% \\ 
  0.7 & 44.9\% & 6.8\% & 9.4\% & 27.6\% & 11.3\% &  & 41.6\% & 11.0\% & 6.7\% & 22.6\% & 18.1\% \\ 
  0.8 & 47.1\% & 8.5\% & 8.7\% & 18.7\% & 17.0\% &  & 42.4\% & 9.2\% & 8.4\% & 19.4\% & 20.6\% \\ 
  0.9 & 57.1\% & 5.4\% & 7.3\% & 7.1\% & 23.1\% &  & 44.9\% & 8.4\% & 8.9\% & 11.2\% & 26.6\% \\ 
\multicolumn{12}{c}{\underline{Loss to CSA}}\\
  0.1 & NA & 74.4\% & 82.9\% & 55.8\% & 75.9\% &  & NA & 73.8\% & 84.4\% & 61.2\% & 70.0\% \\ 
  0.2 & NA & 75.5\% & 83.2\% & 59.4\% & 79.0\% &  & NA & 73.1\% & 83.3\% & 60.4\% & 70.8\% \\ 
  0.3 & NA & 79.7\% & 84.0\% & 59.8\% & 80.3\% &  & NA & 76.3\% & 83.8\% & 58.7\% & 69.9\% \\ 
  0.4 & NA & 80.8\% & 84.2\% & 57.2\% & 76.3\% &  & NA & 76.2\% & 83.9\% & 62.5\% & 69.7\% \\ 
  0.5 & NA & 82.6\% & 84.6\% & 58.9\% & 77.4\% &  & NA & 75.0\% & 82.6\% & 60.7\% & 68.8\% \\ 
  0.6 & NA & 81.8\% & 84.6\% & 60.7\% & 74.6\% &  & NA & 78.9\% & 83.5\% & 63.5\% & 70.2\% \\ 
  0.7 & NA & 83.3\% & 83.6\% & 62.4\% & 75.8\% &  & NA & 80.5\% & 86.9\% & 62.4\% & 66.9\% \\ 
  0.8 & NA & 83.1\% & 84.8\% & 67.9\% & 74.5\% &  & NA & 81.5\% & 83.9\% & 65.8\% & 66.8\% \\ 
  0.9 & NA & 88.3\% & 88.5\% & 83.7\% & 72.0\% &  & NA & 84.1\% & 85.0\% & 76.3\% & 64.8\% \\ 
\hline
\end{tabular}
\end{table}

\begin{figure}[tbp]
\caption{Prediction Errors over Various Quantiles}
\label{fig:quantiles}\vskip10pt \centering
\resizebox{15cm}{!}{
\begin{tabular}{cc}
$n=50$ & $n=150$ \\
\includegraphics[scale=0.6]{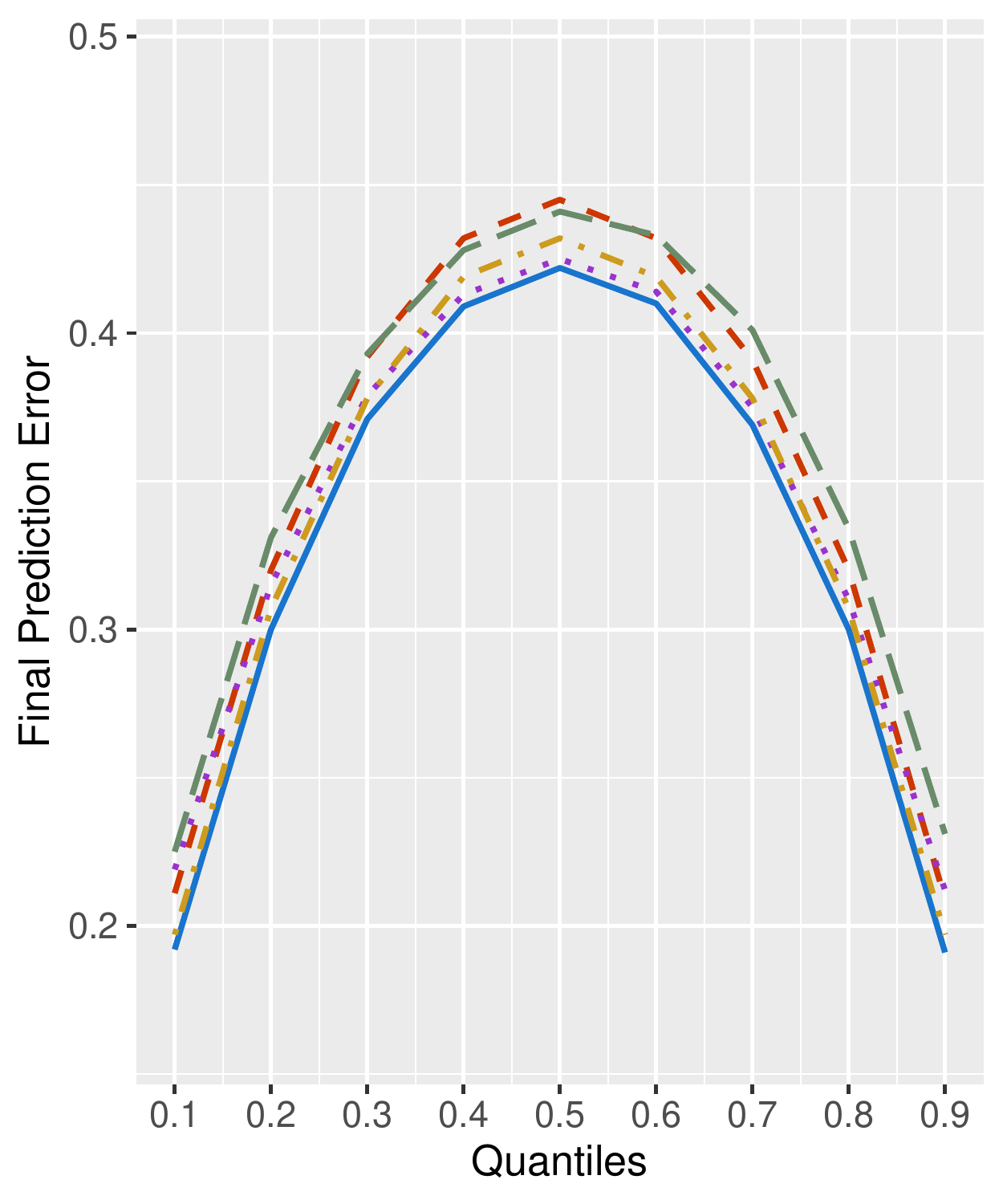} & %
\includegraphics[scale=0.6]{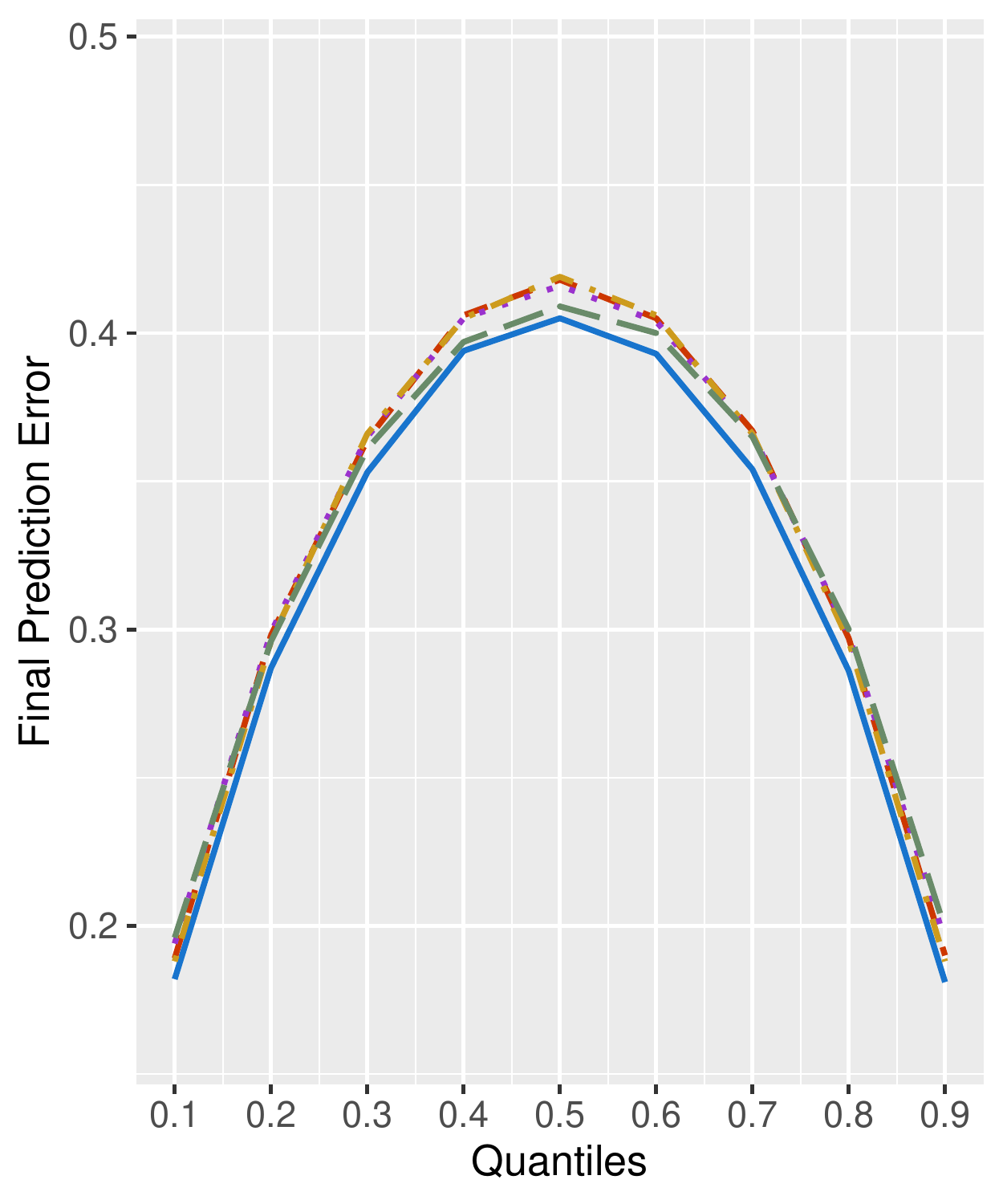}%
\end{tabular}%
}
\end{figure}

\begin{figure}[tbph!]
\caption{Prediction Errors over Various $\protect\rho_x$}
\label{fig:rhox}\vskip10pt \centering
\resizebox{15cm}{!}{
\begin{tabular}{cc}
$n=50$ & $n=150$ \\
\includegraphics[scale=0.6]{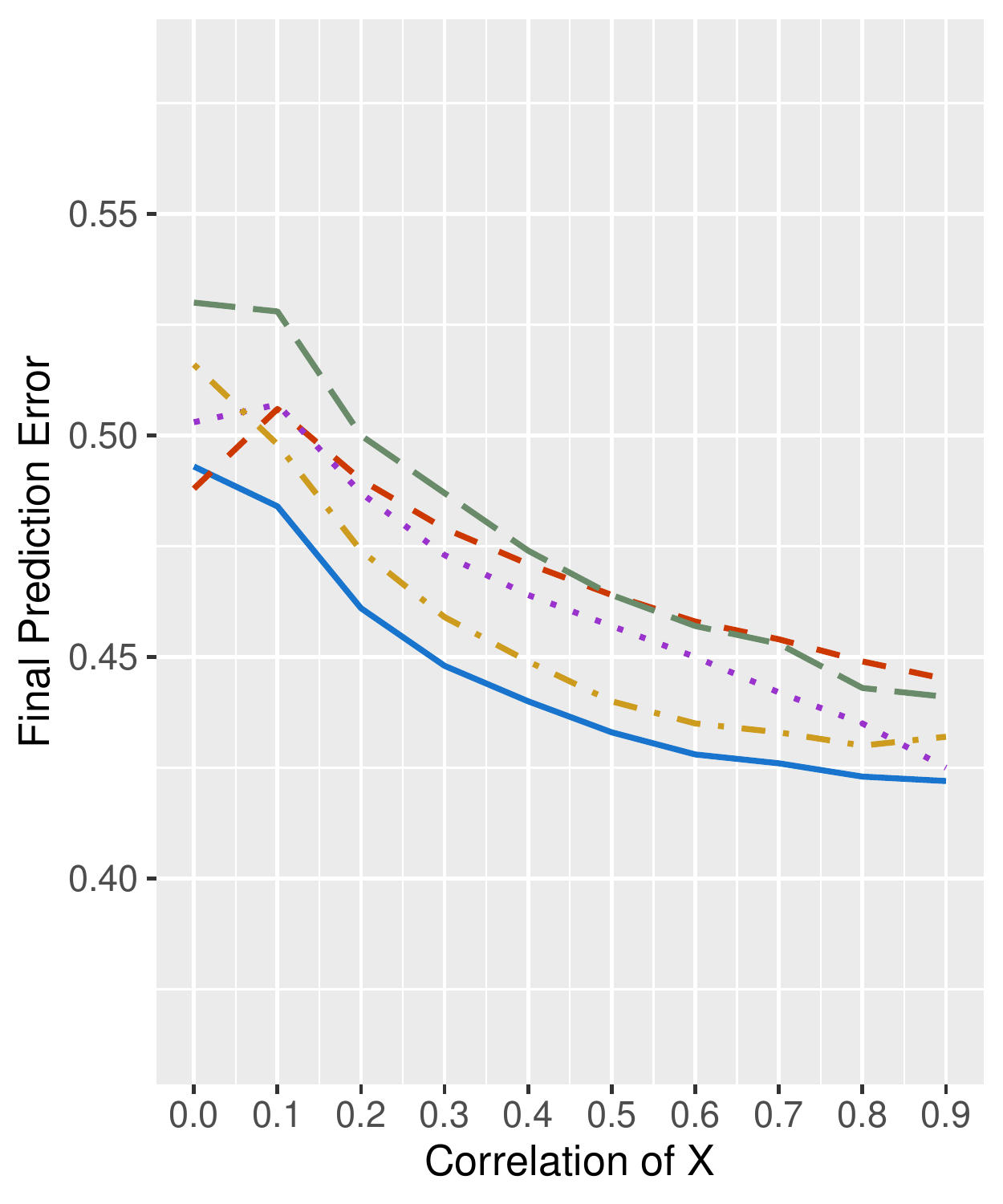} & %
\includegraphics[scale=0.6]{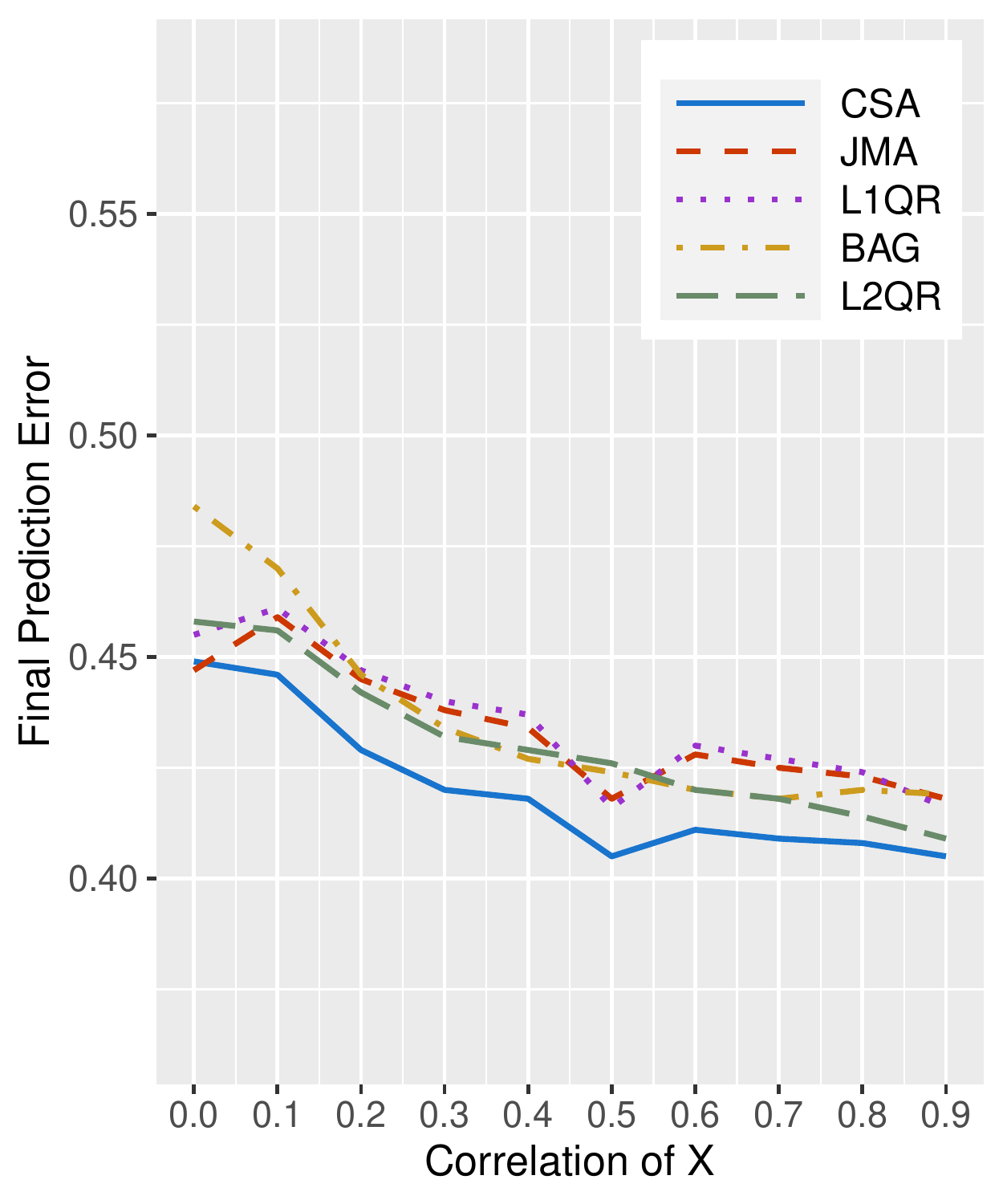}%
\end{tabular}
}
\end{figure}

\begin{table}[tbp]
\centering
\caption{Simulation Results over Various Quantiles}\label{tb:quant}
\begin{tabular}{lccccccccccc}
\\
\hline
      &  \multicolumn{5}{c}{$n=50$}    &  & \multicolumn{5}{c}{$n=150$}   \\
\cline{2-6} \cline{8-12}
$\tau$ &  CSA  & JMA & L1QR & BAG & L2QR & & CSA  & JMA & L1QR & BAG & L2QR \\
\hline
\multicolumn{12}{c}{\underline{Average FPE}}\\
  \multirow{2}{*}{0.1} 	&  0.192 & 0.211 & 0.219 & 0.197 & 0.225 &  & 0.182 & 0.189 & 0.194 & 0.188 & 0.196 \\ 
      					&  (0.029) & (0.036) & (0.040) & (0.023) & (0.046) &  & (0.019) & (0.021) & (0.025) & (0.020) & (0.026) \\ 
  \multirow{2}{*}{0.2}	&  0.300 & 0.320 & 0.315 & 0.307 & 0.331 &  & 0.287 & 0.298 & 0.299 & 0.296 & 0.296 \\ 
      					&  (0.038) & (0.044) & (0.039) & (0.029) & (0.053) &  & (0.024) & (0.027) & (0.028) & (0.026) & (0.030) \\ 
  \multirow{2}{*}{0.3}	&  0.371 & 0.392 & 0.379 & 0.378 & 0.393 &  & 0.353 & 0.364 & 0.365 & 0.366 & 0.361 \\ 
   	  					&  (0.038) & (0.043) & (0.037) & (0.033) & (0.051) &  & (0.029) & (0.032) & (0.031) & (0.030) & (0.032) \\ 
  \multirow{2}{*}{0.4}	&  0.409 & 0.432 & 0.413 & 0.419 & 0.428 &  & 0.394 & 0.406 & 0.405 & 0.405 & 0.397 \\ 
      					&  (0.040) & (0.044) & (0.036) & (0.035) & (0.053) &  & (0.031) & (0.034) & (0.033) & (0.032) & (0.032) \\ 
  \multirow{2}{*}{0.5}	&  0.422 & 0.445 & 0.425 & 0.432 & 0.441 &  & 0.405 & 0.418 & 0.416 & 0.419 & 0.409 \\ 
      					&  (0.042) & (0.046) & (0.035) & (0.035) & (0.053) &  & (0.032) & (0.034) & (0.033) & (0.032) & (0.032) \\ 
  \multirow{2}{*}{0.6}	&  0.410 & 0.432 & 0.414 & 0.419 & 0.433 &  & 0.393 & 0.405 & 0.404 & 0.406 & 0.400 \\ 
      					&  (0.042) & (0.045) & (0.037) & (0.035) & (0.053) &  & (0.032) & (0.033) & (0.033) & (0.031) & (0.032) \\ 
  \multirow{2}{*}{0.7}	&  0.369 & 0.391 & 0.375 & 0.378 & 0.401 &  & 0.354 & 0.367 & 0.366 & 0.366 & 0.365 \\ 
      					&  (0.041) & (0.045) & (0.036) & (0.033) & (0.053) &  & (0.029) & (0.031) & (0.031) & (0.029) & (0.031) \\ 
  \multirow{2}{*}{0.8}	&  0.300 & 0.320 & 0.310 & 0.307 & 0.334 &  & 0.286 & 0.297 & 0.298 & 0.296 & 0.300 \\ 
      					&  (0.036) & (0.043) & (0.035) & (0.030) & (0.052) &  & (0.025) & (0.027) & (0.027) & (0.026) & (0.029) \\ 
  \multirow{2}{*}{0.9}	&  0.191 & 0.209 & 0.212 & 0.197 & 0.231 &  & 0.181 & 0.190 & 0.195 & 0.188 & 0.199 \\ 
      					&  (0.029) & (0.034) & (0.036) & (0.024) & (0.047) &  & (0.019) & (0.022) & (0.025) & (0.020) & (0.027) \\ 
\multicolumn{12}{c}{\underline{Winning Ratio}}\\
  0.1 & 43.7\% & 7.7\% & 8.9\% & 31.5\% & 8.2\% &  & 38.6\% & 10.7\% & 9.2\% & 26.7\% & 14.8\% \\ 
  0.2 & 41.9\% & 7.1\% & 11.9\% & 25.0\% & 14.1\% &  & 38.5\% & 10.2\% & 7.8\% & 18.7\% & 24.8\% \\ 
  0.3 & 37.0\% & 6.0\% & 13.9\% & 23.3\% & 19.8\% &  & 37.0\% & 10.3\% & 8.7\% & 15.2\% & 28.8\% \\ 
  0.4 & 34.0\% & 5.0\% & 15.7\% & 22.0\% & 23.3\% &  & 35.9\% & 7.8\% & 8.3\% & 12.3\% & 35.7\% \\ 
  0.5 & 35.7\% & 5.9\% & 16.3\% & 18.1\% & 24.0\% &  & 38.3\% & 8.8\% & 7.2\% & 13.5\% & 32.2\% \\ 
  0.6 & 36.5\% & 6.6\% & 15.2\% & 23.9\% & 17.8\% &  & 38.3\% & 8.4\% & 8.0\% & 15.6\% & 29.7\% \\ 
  0.7 & 38.1\% & 8.1\% & 14.7\% & 25.9\% & 13.2\% &  & 42.3\% & 8.5\% & 7.3\% & 18.5\% & 23.4\% \\ 
  0.8 & 41.0\% & 7.2\% & 11.6\% & 28.1\% & 12.1\% &  & 40.5\% & 9.7\% & 8.0\% & 23.2\% & 18.6\% \\ 
  0.9 & 43.3\% & 8.6\% & 9.6\% & 30.7\% & 7.8\% &  & 41.5\% & 11.6\% & 7.6\% & 26.5\% & 12.8\% \\ 
\multicolumn{12}{c}{\underline{Loss to CSA}}\\
  0.1 & NA & 82.6\% & 84.6\% & 58.9\% & 77.4\% &  & NA & 75.0\% & 82.6\% & 60.7\% & 68.8\% \\ 
  0.2 & NA & 82.1\% & 78.0\% & 63.3\% & 71.6\% &  & NA & 78.3\% & 83.2\% & 62.5\% & 61.6\% \\ 
  0.3 & NA & 81.6\% & 70.8\% & 59.4\% & 64.0\% &  & NA & 79.9\% & 83.9\% & 63.2\% & 58.4\% \\ 
  0.4 & NA & 81.4\% & 65.7\% & 58.2\% & 59.4\% &  & NA & 81.9\% & 80.0\% & 62.9\% & 53.0\% \\ 
  0.5 & NA & 80.7\% & 65.6\% & 61.1\% & 62.1\% &  & NA & 81.9\% & 81.8\% & 65.6\% & 55.3\% \\ 
  0.6 & NA & 80.1\% & 67.9\% & 59.8\% & 66.9\% &  & NA & 80.2\% & 80.5\% & 63.1\% & 56.9\% \\ 
  0.7 & NA & 78.9\% & 71.4\% & 60.9\% & 71.1\% &  & NA & 82.5\% & 84.1\% & 64.4\% & 62.3\% \\ 
  0.8 & NA & 82.6\% & 74.3\% & 60.1\% & 72.5\% &  & NA & 81.5\% & 82.1\% & 63.4\% & 65.3\% \\ 
  0.9 & NA & 80.1\% & 81.6\% & 61.2\% & 81.0\% &  & NA & 77.1\% & 85.2\% & 62.2\% & 73.1\% \\ 
\hline
\end{tabular}
\end{table}

\begin{table}[tbp]
\small
\centering
\caption{Simulation Results over Various $\protect\rho_x$}\label{tb:rhox}
\begin{tabular}{lccccccccccc}
\\
\hline
      &  \multicolumn{5}{c}{$n=50$}    &  & \multicolumn{5}{c}{$n=150$}   \\
\cline{2-6} \cline{8-12}
$\rho_x$ &  CSA  & JMA & L1QR & BAG & L2QR & & CSA  & JMA & L1QR & BAG & L2QR \\
\hline
\multicolumn{12}{c}{\underline{Average FPE}}\\
  \multirow{2}{*}{0.0} 	&  0.493 & 0.488 & 0.503 & 0.516 & 0.530 &  & 0.449 & 0.447 & 0.455 & 0.484 & 0.458 \\ 
    					&  (0.047) & (0.050) & (0.052) & (0.042) & (0.065) &  & (0.037) & (0.036) & (0.037) & (0.038) & (0.038) \\ 
  \multirow{2}{*}{0.1} 	&  0.484 & 0.506 & 0.507 & 0.498 & 0.528 &  & 0.446 & 0.459 & 0.461 & 0.470 & 0.456 \\ 
      					&  (0.046) & (0.049) & (0.049) & (0.040) & (0.060) &  & (0.036) & (0.038) & (0.037) & (0.037) & (0.038) \\ 
  \multirow{2}{*}{0.2}	&  0.461 & 0.490 & 0.487 & 0.474 & 0.500 &  & 0.429 & 0.445 & 0.447 & 0.446 & 0.442 \\ 
      					&  (0.045) & (0.049) & (0.045) & (0.039) & (0.053) &  & (0.034) & (0.036) & (0.037) & (0.035) & (0.036) \\ 
  \multirow{2}{*}{0.3}	&  0.448 & 0.479 & 0.473 & 0.459 & 0.487 &  & 0.420 & 0.438 & 0.440 & 0.434 & 0.432 \\ 
   	  					&  (0.044) & (0.049) & (0.042) & (0.037) & (0.054) &  & (0.034) & (0.035) & (0.036) & (0.032) & (0.035) \\ 
  \multirow{2}{*}{0.4}	&  0.440 & 0.471 & 0.464 & 0.449 & 0.474 &  & 0.418 & 0.434 & 0.437 & 0.427 & 0.429 \\ 
      					&  (0.042) & (0.046) & (0.042) & (0.036) & (0.052) &  & (0.032) & (0.034) & (0.034) & (0.032) & (0.034) \\ 
  \multirow{2}{*}{0.5}	&  0.433 & 0.464 & 0.457 & 0.440 & 0.464 &  & 0.405 & 0.418 & 0.416 & 0.424 & 0.426 \\ 
      					&  (0.043) & (0.047) & (0.042) & (0.035) & (0.053) &  & (0.032) & (0.034) & (0.033) & (0.033) & (0.035) \\ 
  \multirow{2}{*}{0.6}	&  0.428 & 0.458 & 0.450 & 0.435 & 0.457 &  & 0.411 & 0.428 & 0.430 & 0.420 & 0.420 \\ 
      					&  (0.042) & (0.047) & (0.041) & (0.033) & (0.053) &  & (0.032) & (0.034) & (0.034) & (0.032) & (0.034) \\ 
  \multirow{2}{*}{0.7}	&  0.426 & 0.454 & 0.442 & 0.433 & 0.453 &  & 0.409 & 0.425 & 0.427 & 0.418 & 0.418 \\ 
      					&  (0.041) & (0.046) & (0.038) & (0.034) & (0.053) &  & (0.033) & (0.033) & (0.034) & (0.032) & (0.033) \\ 
  \multirow{2}{*}{0.8}	&  0.423 & 0.449 & 0.435 & 0.430 & 0.443 &  & 0.408 & 0.423 & 0.424 & 0.420 & 0.414 \\ 
      					&  (0.040) & (0.046) & (0.038) & (0.034) & (0.052) &  & (0.031) & (0.034) & (0.033) & (0.033) & (0.034) \\ 
  \multirow{2}{*}{0.9}	&  0.422 & 0.445 & 0.425 & 0.432 & 0.441 &  & 0.405 & 0.418 & 0.416 & 0.419 & 0.409 \\ 
      					&  (0.042) & (0.046) & (0.035) & (0.035) & (0.053) &  & (0.032) & (0.034) & (0.033) & (0.032) & (0.032) \\ 
\multicolumn{12}{c}{\underline{Winning Ratio}}\\
  0.0 & 20.2\% & 33.9\% & 14.3\% & 15.6\% & 16.0\% &  & 23.0\% & 31.1\% & 8.7\% & 6.7\% & 30.5\% \\ 
  0.1 & 38.4\% & 11.9\% & 10.5\% & 26.6\% & 12.6\% &  & 40.0\% & 11.5\% & 6.4\% & 11.6\% & 30.5\% \\ 
  0.2 & 44.3\% & 7.7\% & 7.9\% & 28.4\% & 11.7\% &  & 43.4\% & 9.1\% & 6.9\% & 16.2\% & 24.4\% \\ 
  0.3 & 45.4\% & 6.4\% & 6.5\% & 29.4\% & 12.3\% &  & 44.5\% & 7.1\% & 5.4\% & 19.1\% & 23.9\% \\ 
  0.4 & 46.0\% & 6.6\% & 6.2\% & 28.6\% & 12.6\% &  & 43.4\% & 8.1\% & 5.4\% & 21.2\% & 21.9\% \\ 
  0.5 & 42.5\% & 5.3\% & 7.1\% & 31.7\% & 13.4\% &  & 46.6\% & 10.2\% & 8.6\% & 18.3\% & 16.3\% \\ 
  0.6 & 44.4\% & 4.8\% & 6.3\% & 27.1\% & 17.4\% &  & 43.7\% & 5.8\% & 4.2\% & 21.4\% & 24.9\% \\ 
  0.7 & 41.1\% & 6.8\% & 7.7\% & 27.0\% & 17.4\% &  & 44.5\% & 5.7\% & 3.3\% & 20.1\% & 26.4\% \\ 
  0.8 & 38.7\% & 6.7\% & 10.3\% & 22.3\% & 22.0\% &  & 38.7\% & 7.9\% & 4.3\% & 17.5\% & 31.6\% \\ 
  0.9 & 35.8\% & 5.9\% & 16.3\% & 18.0\% & 24.0\% &  & 38.5\% & 8.7\% & 7.2\% & 13.5\% & 32.1\% \\ 
\multicolumn{12}{c}{\underline{Loss to CSA}}\\
  0.0 & NA & 40.4\% & 63.0\% & 63.9\% & 68.3\% &  & NA & 44.0\% & 67.7\% & 73.5\% & 58.0\% \\ 
  0.1 & NA & 73.7\% & 77.0\% & 61.5\% & 73.7\% &  & NA & 76.6\% & 82.7\% & 70.1\% & 59.0\% \\ 
  0.2 & NA & 79.1\% & 81.0\% & 61.3\% & 73.9\% &  & NA & 81.4\% & 86.5\% & 65.7\% & 62.9\% \\ 
  0.3 & NA & 83.8\% & 82.9\% & 58.8\% & 73.2\% &  & NA & 84.9\% & 88.7\% & 63.3\% & 61.7\% \\ 
  0.4 & NA & 83.9\% & 82.6\% & 59.5\% & 72.1\% &  & NA & 82.6\% & 87.9\% & 61.4\% & 61.7\% \\ 
  0.5 & NA & 87.1\% & 82.5\% & 55.8\% & 70.2\% &  & NA & 81.9\% & 81.8\% & 67.3\% & 70.2\% \\ 
  0.6 & NA & 85.7\% & 82.7\% & 58.4\% & 68.6\% &  & NA & 85.3\% & 88.6\% & 59.3\% & 59.1\% \\ 
  0.7 & NA & 84.3\% & 79.1\% & 57.4\% & 67.7\% &  & NA & 85.2\% & 88.8\% & 58.8\% & 58.9\% \\ 
  0.8 & NA & 81.3\% & 74.5\% & 59.3\% & 63.1\% &  & NA & 83.3\% & 87.8\% & 61.7\% & 53.9\% \\ 
  0.9 & NA & 80.7\% & 65.6\% & 61.6\% & 62.1\% &  & NA & 81.9\% & 81.8\% & 65.6\% & 55.4\% \\ 
\hline
\end{tabular}

\end{table}

Figures \ref{fig:R2}--\ref{fig:rhox} and Tables \ref{tb:R2-0.5quant}--\ref{tb:rhox} summarize the simulation results over all designs. 
Overall, the performance of CSA compared to the alternative is quite satisfactory.
We first \changed{direct} our attention to Figure \ref{fig:R2} and Tables \ref{tb:R2-0.5quant}--\ref{tb:R2-0.1quant}.
In these simulation designs, we vary $R^2$ over $\{0.1,0.2,\ldots, 0.9\}$ while setting $\rho_x=0.9$.
We consider two quantiles, $\tau=0.1$ and $0.5$, respectively.
From the four graphs in Figure \ref{fig:R2}, we confirm that CSA outperforms the alternative uniformly over $R^2$'s in terms of FPE in both quantiles.
The prediction performance of CSA is better when the sample size is small, $n=50$, and the gap decreases as the
sample size increases to $n=150$.
At $\tau=0.5$, L1QR performs the second when $n=50$ but L2QR does the second when $n=150$. Thus, the performance order next to CSA is not stable. 
AT $\tau=0.1$, BAG performs the second overall but it is deteriorated when $R^2$ is very high, e.g.\ $R^2=0.9$.
We also note that the performance of CSA is relatively stable over $R^2$ while that of the alternative increases steeply for larger $R^2$ when $n=50$.
The same results are confirmed in Tables \ref{tb:R2-0.5quant}--\ref{tb:R2-0.1quant}.
CSA shows the highest winning ratios over all designs except $\tau=0.5$ and $R^2=0.1$, where that of L2QR is slightly higher. 
When we conduct the binary comparison \changed{(loss to CSA}), all methods lose more than 50\% to CSA over all designs \changed{and more than 80\% in some designs.}
Therefore, we conclude that both the winning ratio and the loss to CSA are more favorable to CSA in this set of simulation designs. 

In the next simulation, we study the performance over a wider range of quantiles. We vary the quantile $\tau=\{0.1,0.2,\ldots,0.9\}$ while setting $R^2=0.5$ and $\rho_x=0.9$.
The results are summarized in Figure \ref{fig:quantiles} and Table \ref{tb:quant}.
In Figure \ref{fig:quantiles}, CSA outperforms the alternative uniformly over all quantiles in both sample sizes followed by BAG and L2QR.
Again, the gap decreases as the sample size increases.
It is also interesting that all estimators predict better at the tail distributions and they show the largest prediction errors at the median.
The winning ratio and the loss to CSA in Table \ref{tb:quant} are also satisfactory.

Third, we check the performance over different levels of dependency among the predictors. We vary $\rho_x=\{0,0.1,0.2,\ldots,0.9\}$ while setting $R^2=0.5$ and $\tau=0.5$.
Since $(x_{i2},\ldots,x_{i1000})$ are generated from the multivariate normal distribution, they are independent when $\rho_x=0$.
Figure \ref{fig:rhox} reveals an interesting point. CSA performs better than the alternative when there exists any correlation between the predictors, i.e.\  $\rho_x > 0$.
Recall that most simulation studies in the literature consider independent predictors. 
As we can see from the empirical applications in the next section, however, the predictors are usually correlated with each other.
Therefore, it is promising that CSA performs better when there is any correlation among predictors.
\citet*{elliott2013complete} also report in the conditional mean prediction settings that the CSA approach performs better when predictors are correlated \changed{with} each other.
In Table \ref{tb:rhox}, both the winning ratio and the loss to CSA statistics improve dramatically when $\rho_x$ is away from zero, where JMA performs the best.

\begin{table}[tbp]
\centering
\caption{Correct Specification: Decreasing Signal}\label{tb:correct-decreasing}
\begin{tabular}{lccccc}
\\
\hline
\multicolumn{6}{c}{$n=50$}\\
\cline{2-6}
   & CSA   &  JMA   &  L1QR  & BAG & L2QR \\
\hline
\multicolumn{6}{l}{\underline{Average FPE}}\\
\multirow{2}{*}{$K=5$}   	&  0.415 & 0.424 & 0.419 & 0.430 & 0.419 \\
    						&  (0.033) & (0.036) & (0.034) & (0.036) & (0.035)  \\ 
\multirow{2}{*}{$K=15$}   	&  0.420 & 0.443 & 0.424 & 0.426 & 0.427  \\ 
    						&  (0.039) & (0.044) & (0.034) & (0.035) & (0.047) \\ 
\multicolumn{6}{l}{\underline{Winning Ratio}}\\
$K=5$   &  28.5\% & 11.1\% & 13.2\% & 12.5\% & 34.7\%   \\ 
$K=15$  &  32.7\% & 5.4\% & 11.1\% & 21.7\% & 29.1\%    \\ 
\multicolumn{6}{l}{\underline{Loss to CSA}}\\
$K=5$   &  NA & 73.6\% & 66.2\% & 62.7\% & 53.7\% \\ 
$K=15$  &  NA & 81.8\% & 70.1\% & 56.7\% & 54.4\% \\ 
\hline
\multicolumn{6}{c}{$n=150$}\\
\cline{2-6}
   & CSA   &  JMA   &  L1QR  & BAG & L2QR \\
\hline
\multicolumn{6}{l}{\underline{Average FPE}}\\
\multirow{2}{*}{$K=10$}   	&  0.406 & 0.412 & 0.411 & 0.422 & 0.406 \\ 
    						&  (0.032) & (0.033) & (0.032) & (0.034) & (0.031) \\ 
\multirow{2}{*}{$K=20$}   	&  0.407 & 0.419 & 0.419 & 0.417 & 0.408 \\ 
    						&  (0.032) & (0.034) & (0.033) & (0.032) & (0.032) \\ 
\multicolumn{6}{l}{\underline{Winning Ratio}}\\
$K=10$  &  30.6\% & 10.8\% & 8.6\% & 8.7\% & 41.3\% \\ 
$K=20$  &  35.2\% & 9.2\% & 6.5\% & 13.0\% & 36.1\% \\ 
\multicolumn{6}{l}{\underline{Loss to CSA}}\\
$K=10$  &  NA & 71.6\% & 72.7\% & 65.0\% & 49.0\% \\ 
$K=20$  &  NA & 77.9\% & 82.9\% & 61.6\% & 50.9\% \\ 
\hline
\end{tabular}
\end{table}

\begin{table}[tbp]
\centering
\caption{Correct Specification: Constant Signal}\label{tb:correct-constant}
\begin{tabular}{lccccc}
\\
\hline
\multicolumn{6}{c}{$n=50$}\\
\cline{2-6}
   & CSA   &  JMA   &  L1QR  & BAG & L2QR \\
\hline
\multicolumn{6}{l}{\underline{Average FPE}}\\
\multirow{2}{*}{$K=5$}   	&  0.414 & 0.426 & 0.419 & 0.429 & 0.415 \\ 
    						&  (0.033) & (0.036) & (0.034) & (0.035) & (0.034) \\ 
\multirow{2}{*}{$K=15$}   	&  0.418 & 0.443 & 0.422 & 0.429 & 0.429 \\ 
    						&  (0.040) & (0.046) & (0.036) & (0.034) & (0.049) \\ 
\multicolumn{6}{l}{\underline{Winning Ratio}}\\
$K=5$   &  30.3\% & 7.0\% & 12.6\% & 12.4\% & 37.7\% \\ 
$K=15$  &  33.1\% & 4.7\% & 14.5\% & 16.9\% & 30.8\% \\ 
\multicolumn{6}{l}{\underline{Loss to CSA}}\\
$K=5$   &  NA & 80.6\% & 68.0\% & 61.6\% & 50.6\% \\ 
$K=15$  &  NA & 85.0\% & 65.8\% & 60.2\% & 56.0\% \\ 
\hline
\multicolumn{6}{c}{$n=150$}\\
\cline{2-6}
   & CSA   &  JMA   &  L1QR  & BAG & L2QR \\
\hline
\multicolumn{6}{l}{\underline{Average FPE}}\\
\multirow{2}{*}{$K=10$}   	&  0.406 & 0.414 & 0.411 & 0.422 & 0.406 \\ 
    						&  (0.031) & (0.032) & (0.032) & (0.032) & (0.032) \\ 
\multirow{2}{*}{$K=20$}   	&  0.406 & 0.419 & 0.416 & 0.420 & 0.410 \\ 
    						&  (0.033) & (0.034) & (0.034) & (0.033) & (0.032) \\ 
\multicolumn{6}{l}{\underline{Winning Ratio}}\\
$K=10$  &  31.2\% & 6.2\% & 11.4\% & 8.4\% & 42.8\% \\ 
$K=20$  &  40.3\% & 6.6\% & 8.8\% & 11.9\% & 32.4\% \\ 
\multicolumn{6}{l}{\underline{Loss to CSA}}\\
$K=10$  &  NA & 78.3\% & 73.7\% & 67.4\% & 47.8\% \\ 
$K=20$  &  NA & 82.2\% & 81.8\% & 65.3\% & 56.3\% \\ 
\hline
\end{tabular}
\end{table}

\begin{table}[tbp]
\centering
\caption{Correct Specification: Sparse Signal}\label{tb:correct-sparse}
\begin{tabular}{lccccc}
\\
\hline
\multicolumn{6}{c}{$n=50$}\\
\cline{2-6}
   & CSA   &  JMA   &  L1QR  & BAG & L2QR \\
\hline
\multicolumn{6}{l}{\underline{Average FPE}}\\
\multirow{2}{*}{$K=5$}   	&  0.422 & 0.422 & 0.420 & 0.430 & 0.421 \\ 
    						&  (0.034) & (0.036) & (0.035) & (0.036) & (0.035) \\ 
\multirow{2}{*}{$K=15$}   	&  0.437 & 0.442 & 0.431 & 0.430 & 0.439 \\ 
    						&  (0.042) & (0.046) & (0.038) & (0.035) & (0.049) \\ 
\multicolumn{6}{l}{\underline{Winning Ratio}}\\
$K=5$   &  13.5\% & 20.9\% & 20.3\% & 12.7\% & 32.6\% \\ 
$K=15$  &  14.7\% & 16.9\% & 18.3\% & 25.6\% & 24.5\% \\ 
\multicolumn{6}{l}{\underline{Loss to CSA}}\\
$K=5$   &  NA & 47.2\% & 41.9\% & 57.2\% & 50.4\% \\ 
$K=15$  &  NA & 52.3\% & 44.9\% & 45.5\% & 48.8\% \\ 
\hline
\multicolumn{6}{c}{$n=150$}\\
\cline{2-6}
   & CSA   &  JMA   &  L1QR  & BAG & L2QR \\
\hline
\multicolumn{6}{l}{\underline{Average FPE}}\\
\multirow{2}{*}{$K=10$}   	&  0.414 & 0.410 & 0.411 & 0.423 & 0.411 \\ 
    						&  (0.032) & (0.032) & (0.032) & (0.032) & (0.031) \\ 
\multirow{2}{*}{$K=20$}   	&  0.418 & 0.416 & 0.419 & 0.422 & 0.415 \\ 
    						&  (0.033) & (0.034) & (0.034) & (0.031) & (0.032) \\ 
\multicolumn{6}{l}{\underline{Winning Ratio}}\\
$K=10$  &  13.3\% & 22.8\% & 15.5\% & 12.3\% & 36.1\% \\ 
$K=20$  &  13.9\% & 24.9\% & 12.6\% & 17.4\% & 31.2\% \\ 
\multicolumn{6}{l}{\underline{Loss to CSA}}\\
$K=10$  &  NA & 38.2\% & 39.5\% & 58.0\% & 45.2\% \\ 
$K=20$  &  NA & 41.0\% & 51.0\% & 52.3\% & 45.9\% \\ 
\hline
\end{tabular}
\end{table}

We next consider the second category of simulation designs, where the candidate models include the true DGP.
The new simulations are based on the following model:
\eqs{
  y_i = \theta \sum_{j=1}^{K} \beta_{j} x_{ij} + \varepsilon _i, \label{eq:sim design2}
}where we observe all $K$ predictors in the sample. We consider $K=5,15$ when $n=50$ and $K=10,20$ when $n=150$. Similar to the previous simulations, the population $R^2$ is controlled by $\tht$. 
We set $R^2=0.5$, $\tau=0.5$, and $\rho_x=0.9$. Instead of varying $R^2$, $\tau$, and $\rho_X$, we consider three signal structures in this simulation: 
\eqs{
	\mbox{Decreasing signal :}~~ & \beta_j=j^{-1} \\	
	\mbox{Constant signal :}~~ & \beta_j = 1 \mbox{ for all } j\\
	\mbox{Sparse signal :}~~ & \beta_j=\begin{cases} 1 \mbox{ if } j=1,2 \\ 0 \mbox{ if } j > 2\end{cases}. 
}
Therefore, we consider 12 new DGPs in total.

Tables \ref{tb:correct-decreasing}-\ref{tb:correct-sparse} summarize the simulation results. First of all, we take a look at the loss to CSA ratio in the second column (JMA) in these tables. 
Note that the loss ratio increases as $K$ increases over all different designs, which is expected by the theoretical results in Theorem \ref{thm:prediction-efficiency-bound}.
Second, CSA performs worse in the sparse signal models compared to the other two designs. As discussed under equation \eqref{eq:obj-mean-expression}, this is expected from the theory in Section \ref{sc: asympt} since the sparse design generates many subsets with totally irrelevant predictors.
Third, it is interesting that JMA does not particularly \changed{outperform} in this setup, where the candidate models include the true one. Also, note that L1QR does not particularly outperform in the sparse signal model.
In fact, L2QR performs well over all three signal designs. Given that L2QR is understudied in the literature, this would be an interesting topic for future research.

In sum, we confirm that CSA shows satisfactory finite sample properties via Monte Carlo simulation studies. Related to the forecast combination puzzle, we observe a similar phenomenon in quantile forecasting and confirm some theoretical predictions developed in Section \ref{sc: asympt}.

\section{Empirical Illustration}\label{sc:empirical}

In this section, we investigate the performance of the proposed method with real data sets. Specifically, we revisit two empirical applications in \citet{lu2015jackknife}: (i) quantile forecast of excess stock returns; and (ii) quantile forecast of wages. Following the simulation studies in Section \ref{sc:simulations}, we compare the performance of the complete subset averaging (CSA) method to the Jackknife Model Averaging (JMA), the $\ell_1$-penalized quantile regression (L1QR), the bootstrap aggregating method (BAG), and the $\ell_2$-penalized quantile regression (L2QR).

\begin{table}[tbp]
\caption{Out-of-sample $R^2$ for the Excess Stock Return Data}
\label{tb:stock}
\centering
\vskip10pt
\begin{threeparttable}
\begin{tabular}{ccrrrrrrrrr}
\hline
$\tau$ & $T_1$ & & \multicolumn{1}{c}{CSA} & \multicolumn{1}{c}{JMA} & \multicolumn{1}{c}{L1QR} & \multicolumn{1}{c}{BAG} & \multicolumn{1}{c}{L2QR} & & $E[\widehat {k}]$ & $Med[\widehat {k}]$ \\
\hline
\\
\multirow{7}{*}{0.05} 	& 48 & & -0.071 (2)  &-0.117 (4) & -0.088 (3)  & 0.031 (1)  & -4.331  (5) &  &7.4&8 \\
						& 60 & & -0.063 (3)  &-0.125 (4) & -0.038 (2)  & 0.009 (1)  & -4.395  (5) & &8.1&8 \\
						& 72 & & -0.001 (2)  &-0.023 (4) & -0.005 (3)  & 0.020 (1)  & -3.955  (5) & &8.2&9 \\
						& 96 & & 0.055  (1)  &-0.020 (4) & -0.012 (3)  & 0.027 (2)  & -4.010  (5) & &8.1&9 \\
						& 120& & 0.104  (1)  &0.053  (2) & 0.028  (4) & 0.033  (3) & -3.655   (5) & &8.7&9 \\
						& 144& & 0.082  (1)  &0.045  (2) & 0.012  (4) & 0.019  (3) & -3.735   (5) & &9.1&9 \\
						& 180& & 0.039  (1)  &0.033  (2) & 0.023  (3) & -0.011 (4)  & -2.311  (5) & &9.6&10 \\
\\
\multirow{7}{*}{0.5}	& 48  & &0.103	(1) & 0.076  (2)  & -0.040 (4) & -0.016 (3) & -2.341 (5) & &9.8&10 \\
						& 60  & &0.089	(1) & 0.079	(2)  & -0.036 (4) & -0.013 (3) & -2.078 (5) & &9.9&10 \\
						& 72  & &0.057	(2) & 0.067	(1)  & -0.003 (3) & -0.009 (4) & -1.953 (5) & &10.0&10 \\
						& 96  & &0.049	(2) & 0.053	(1)  & -0.013 (3) & -0.014 (4) & -2.206 (5) & &10.3&11 \\
						& 120 & &0.003	(2) & 0.013	(1)  & 0.003  (3) & -0.011 (4) & -1.882 (5) & &10.5&11 \\
						& 144 & &-0.012 (3) & -0.002 (1) & -0.006 (2) & -0.022 (4) & -1.648 (5) & &10.6&11 \\
						& 180 & &0.032	(2) & 0.034	(1)  & 0.018  (3) & -0.012 (4) & -1.031 (5) & &10.5&11 \\
\\
\hline
\end{tabular}
    \begin{tablenotes}
      \small
      \item Notes: The number in the parentheses denotes the performance ranking among the five different methods. 
    \end{tablenotes}
\end{threeparttable}
\end{table}

\subsection{Stock Return}

The same data set is composed of monthly observations of the US stock market from January 1950 to December 2005 ($T=672$). The
dependent variable is the excess stock return. We use the following twelve regressors: default yield spread, treasury bill rate, net equity expansion, term spread, dividend price ratio, earnings price ratio, long term yield, book-to-market ratio, inflation, return on equity, lagged dependent variable, and smoothed earnings price ratio. See \citet{lu2015jackknife} and \citet{campbell2007predicting} for the details of the data set. Note that JMA \changed{needs} to select the order of important regressors, but we do not need such a selection for CSA, BAG, L2QR. L1QR would select important regressors automatically by the $\ell_1$-penalty. 

We forecast the one-period-ahead excess stock returns at 0.5 and 0.05 quantiles using various fixed in-sample sizes, $T_1 = 48, 60, 72, 96, 120, 144,\mbox{ and }180$. The forecast performance is measured by the out-of-sample $R^2$ defined as
\begin{align*}
R^2 = 1 - \frac{\sum_{t=T_1}^{T-1} \rho_{\tau}(y_{t+1}-\widehat {y}_{t+1|t})%
}{\sum_{t=T_1}^{t-1} \rho_{\tau}(y_{t+1}-\bar{y}_{t+1|t})}
,\end{align*}%
where $\widehat {y}_{t+1|t}$ the one-period-ahead $\tau$-quantile prediction at time $t$ using the data from the past $T_1$ periods, and $\bar{y}_{t+1|t}$ is the unconditional $\tau$-quantile for the same $T_1$ periods. The out-of-sample $R^2$ measures the relative performance of a forecast method compared to the unconditional historical quantile. The higher values of $R^2$ imply better forecasting performance.

Table \ref{tb:stock} summarizes the forecasting results. In addition to $R^2$, we report the ranking of each forecasting method, the mean of $\widehat {k}$, and the median of $\widehat {k}$. The upper panel of Table \ref{tb:stock} reports the results when $\tau=0.05$. The $R^2$ of CSA is better than that of JMA \emph{uniformly} over different sample sizes ($T_1$). The gap between two $R^2$'s is substantial except $T=180$. BAG performs well when $T_1$ is small. The performance of L2QR is not satisfactory over all in-sample sizes. 
We next turn our attention to the lower panel when $\tau=0.5$. Again, CSA performs \changed{the best or second best except when} $T_1=144$. CSA performs better when $T_1$ is small while JMA does better when $T_1$ is larger. Overall, the gap between $R^2$'s is small when $\tau=0.5$. As we have observed from the simulation studies, the performance of \changed{the} two estimators becomes similar as the sample size increases in both panels. It is also noticeable that the selected $\widehat{k}$ of CSA increases as the sample size increases and that CSA selects relatively large $\widehat{k}$ across all $T_1$ and $\tau$. Different from $\tau=0.05$, BAG performs poorly when $\tau=0.5$. L2QR also shows poor performance. 

In sum, the performance of CSA is satisfactory in this forecasting exercise. It is quite stable over different in-sample sizes ($T_1$) and different quantiles in terms of the performance ranking. Among the alternative, BAG and JMA perform well in certain quantiles (0.05 and 05, respectively), but they do poorly when we apply them in different quantiles.

\subsection{Wage}

\begin{table}[tbp]
\caption{Out-of-sample $R^2$ for the Wage Data}
\label{tb:wage}\centering
\vskip10pt
\begin{threeparttable}
\begin{tabular}{ccrrrrrrrrr}
\hline
$\tau$ & $n_1$ & & \multicolumn{1}{c}{CSA} & \multicolumn{1}{c}{JMA} & \multicolumn{1}{c}{L1QR} & \multicolumn{1}{c}{BAG} & \multicolumn{1}{c}{L2QR} & & $E[\widehat {k}]$ & $Med[\widehat {k}]$ \\
\hline
\\
\multirow{4}{*}{0.05} 	& 50   & & 0.066 (2)  & -0.034 (3) & -0.035 (4) & 0.104 (1) & -0.139 (5) &  & 3.9 & 4 \\
						& 100  & & 0.122 (2) & 0.073 (4)  & 0.078 (3) & 0.133 (1) & 0.020 (5)  &  & 5.5 & 6 \\
						& 150  & & 0.138 (2) & 0.112 (4)  & 0.113 (3)  & 0.144 (1) & 0.076 (5) &  & 6.1 & 6 \\
						& 200  & & 0.158 (1)  & 0.125 (4)  & 0.132 (3)  & 0.154 (2) & 0.111 (5) &  & 6.6 & 7 \\
\\
\multirow{4}{*}{0.5}	& 50   & & 0.252 (1)  & 0.233 (3)	& 0.198 (5) & 0.248 (2) & 0.212 (4) & & 6.5 & 6 \\
						& 100  & & 0.287 (1)  & 0.276 (3) & 0.233 (5) & 0.285 (2) & 0.260 (4) & & 7.7 & 8 \\
						& 150  & & 0.302 (1)  & 0.293 (3)	& 0.253 (5) & 0.301 (2) & 0.290 (4) & & 8.2 & 8 \\
						& 200  & & 0.307 (2)  & 0.302 (3)	& 0.267 (5) & 0.312 (1) & 0.302 (4) & & 8.4 & 9 \\
\\
\hline
\end{tabular}%
    \begin{tablenotes}
      \small
      \item Notes: The number in the parentheses denotes the performance ranking among the five different methods. 
    \end{tablenotes}
\end{threeparttable}
\end{table}

In this subsection we conduct the quantile forecast exercises using the
Current Population Survey (CPS) data in 1975. The same data set is also used by
\citet{lu2015jackknife} and \citet{hansen2012jackknife} for quantile and mean forecast exercises, respectively. The sample size is $n=526$ and we use the logarithm of the average hourly wage as the dependent variable. We use the following ten regressors: professional occupation, years of education, years with current employer, female, service occupation, married, trade, SMSA, services, and clerk occupation.

We split the sample into the estimation sample randomly drawn $n_1$
observations and the evaluation sample of $n-n_1$ observations. The
estimation sample size varies $n_1=50,100, 150,$ and $200$ and the random
splitting is repeated 200 times for each $n_1$. The out-of-sample $R^2$ is
defined as
\begin{align*}
R^2 = 1 - \frac{\sum_{s=1}^{n_2} \rho_{\tau}(y_s-\widehat {y}_s)}{%
\sum_{s=1}^{n_2} \rho_{\tau}(y_s-\bar{y}_s)},
\end{align*}%
where $\widehat {y}_s$ is the $\tau$-th conditional quantile predictor and $%
\bar{y}_s$ is the unconditional $\tau$-quantile estimate from the estimation
sample. Again, $R^2$ measures the prediction performance relative
to the unconditional quantile estimate. 

Table \ref{tb:wage} summarizes the exercise results.\footnote{$R^2$s of JMA are different from the numbers reported in Table 5 in \citet{lu2015jackknife} because they implemented the level of wage as a dependent variable which is supposed to be $\log(wage)$. We use $\log(wage)$ in this empirical illustration.} 
We confirm that CSA shows good and stable quantile prediction performance. 
In this application, BAG shows quite \changed{a} similar performance to CSA. 
Similar to the stock return application, CSA performs better than BAG when $\tau=0.5$ and BAG does when $\tau=0.05$.
The prediction results of JMA, L1QR, and L2QR are worse than CSA and BAG. 
The performance gaps are larger when the sample size ($n_1$) is small and they narrow as $n_1$ increases. 
As predicted by the theory and also confirmed in the stock return application, the selected $\widehat{k}$ increases as $n_1$ increases.

\section{Conclusion}\label{sc:concl}
In this paper, we propose a novel conditional quantile prediction method based on complete subset averaging of quantile regressions. We show the asymptotic properties of the estimator when the dimension of regressors diverges to infinity as the sample size increases. The size of the complete subset is chosen by the leave-one-out cross-validation method. We prove that the subset size chosen by this method is optimal in the sense that it is asymptotically equivalent to the infeasible optimal size minimizing the final prediction error. The prediction performance in the simulation studies and empirical applications is satisfactory.

We conclude with two potential extensions of the \changed{proposed} method. First, we can think of a different approach in choosing the complete subset size. Recently, \citet{hirano2019analyzing} propose a Laplace cross-validation method, where the tuning parameter of interest is chosen by the pseudo-Bayesian posterior mean\changed{,} and show that it works better than the standard cross-validation method when the risk function is asymmetric. It \changed{would be} interesting to check how it performs in the CSA quantile prediction. Second, it will be useful if one can extend  the results into the time-series data possibly including persistent regressors (e.g.\ \citet{fan2019predictive}). We leave them for future research.

\clearpage
\begin{appendix}
\changed{\section*{Appendix}}

\changed{In this appendix, we provide all necessary lemmas and technical proofs of the main text.}
\begin{lemma}
\label{lem:orders} Let $e_{n}:=(nMK^{2})^{1/4}$. Suppose that $K/\log
(n)=O(1)$. Then, we can show the following rate conditions:

\begin{enumerate}
\item[(i)] $M=O(2^{K})$

\item[(ii)] $K\log M/n=o(1)$

\item[(iii)] $(e_{n}\log M)/n=o(1)$\changed{.}
\end{enumerate}
\end{lemma}

\begin{proof} [Proof of Lemma \ref{lem:orders}]
(i) Recall the dependency of $M$ on $K$ and $k$. By construction, $M_{K,k} = \binom{K}{k}$. Then, the result follows from $2^K = \sum_{k=0}^{K}\binom{K}{k}$.\\
(ii) Note that
\eqs{
	\frac{K \log M}{n}  & = O\lt(\frac{\log n (\log 2^{\log n})}{n} \rt)\\
						& = O\lt(\log 2 \cdot \frac{(\log n)^2}{n}  \rt)=o(1)\changed{.}
}
(iii) Note that
\eqs{
	\frac{e_n \log M}{n} & = \frac{  \lt( n M K^2 \rt)^{1/4} \log M }{n}  \\
	& = \lt( \frac{MK^2 \lt(\log M\rt)^4 }{n^3} \rt)^{1/4} \\
	& = O\lt(\lt( \frac{2^{\log n}}{n}  \rt)^{1/4}\rt) O\lt(\lt( \frac{(\log n)^2}{n}  \rt)^{1/4}\rt) O\lt(\lt( \frac{(\log 2^{\log n})^2}{n}  \rt)^{1/4}\rt)\changed{.}
}It is enough to show that $2^{\log n} / n =o(1)$. Let $c_{1n} = 2^{\log n} / n$. Then,
\eqs{
  \log c_{1n} = \log n (\log 2 - 1) \rarrow -\infty.
}Therefore, $c_{1n}=o(1)$ and the desired result is established.
\end{proof}

\begin{lemma}
\label{lem:x-rate} Suppose that (i) $\sup_{j\geq 1}E[x_{ij}^{2}]<c_{x}$ with
$c_{x}<\infty $, (ii) $E[\mu _{i}^{2}]<\infty $. Then,
\begin{equation*}
\max_{1\leq k\leq K}\max_{1\leq m\leq M}\frac{1}{n}\sum_{i=1}^{n}\left\Vert
x_{i(m,k)}\right\Vert =O_{p}(K^{1/2})\changed{.}
\end{equation*}
\end{lemma}
\begin{proof} [Proof of Lemma \ref{lem:x-rate}]
The triangle inequality implies that
\eqs{
& \max_{1 \le k \le K} \max_{1 \le m \le M} \ave \lt\Vert x_{i(m,k)} \rt\Vert \\
& \le
 \max_{1 \le k \le K} \max_{1 \le m \le M} \ave E\lt\Vert x_{i(m,k)} \rt\Vert +  \max_{1 \le k \le K} \max_{1 \le m \le M} \lt\vert \ave   \lt( \lt\Vert x_{i(m,k)} \rt\Vert - E\lt\Vert x_{i(m,k)} \rt\Vert \rt) \rt\vert \\
& \equiv A_{1} + A_{2}.
}We first investigate $A_1$:
\eqs{
A_1
& =   \max_{1 \le k \le K} \max_{1 \le m \le M} \ave E\lt[x_{i(m,k)}'x_{i(m,k)}   \rt]^{1/2} \\
& \le   \max_{1 \le k \le K} \max_{1 \le m \le M} \ave \lt(E \lt[x_{i(m,k)}'x_{i(m,k)}\rt]   \rt)^{1/2}\\
& \le   \max_{1 \le k \le K} \max_{1 \le m \le M} \ave \lt(k  c_x   \rt)^{1/2} \\
& \le   \max_{1 \le k \le K} k^{1/2} c_x^{1/2} \\
& \le   K^{1/2} c_x^{1/2} = O(K^{1/2})\changed{.}
}
We next turn our attention to $A_2$. Let $v_{i(m,k)}:= \lt\Vert x_{i(m,k)} \rt\Vert - E\lt\Vert x_{i(m,k)} \rt\Vert $. Note that $Var(v_{i(m,k)} ) \le C K$ for some generic constant $C>0$. Let $e_n:=(n M K^2)^{1/ 4}$. We have
\eqs{
  P\lt( A_2 \ge 2\eps \rt) & = P\lt( \max_{1 \le k \le K} \max_{1 \le m \le M} \lt\vert \ave  v_{i(m,k)} \rt\vert \ge 2\eps \rt) \\
  & \le P\lt( \max_{1 \le k \le K} \max_{1 \le m \le M}  \ave  \lt\vert v_{i(m,k)} \rt\vert \ge 2\eps \rt) \\
  & \le P\lt( \max_{1 \le k \le K} \max_{1 \le m \le M}  \ave  \lt\vert v_{i(m,k)} \rt\vert 1\lt(\lt\vert v_{i(m,k)} \rt\vert  \le e_n \rt)  \ge \eps \rt) \\
  & \hskip15pt + P\lt( \max_{1 \le k \le K} \max_{1 \le m \le M}  \ave  \lt\vert v_{i(m,k)} \rt\vert 1\lt(\lt\vert v_{i(m,k)} \rt\vert  > e_n \rt)  \ge \eps \rt) \\
  & \equiv A_{21} + A_{22}.
}Boole's and Bernstein inequalities imply that
\eqs{
  A_{21}
  & \le K M \max_{1\le k \le K} \max_{1 \le m \le M} P\lt( \ave  \lt\vert v_{i(m,k)} \rt\vert 1\lt(\lt\vert v_{i(m,k)} \rt\vert  \le e_n \rt)  \ge \eps \rt) \\
  & \le 2 K M \exp\lt( - \frac{n \eps^2}{2CK + 2\eps e_n /3} \rt) \\
  & = 2 \exp\lt( - \frac{n \eps^2}{2CK + 2\eps e_n /3} + \log M + \log K \rt) \\
  & = 2 \exp\lt( \frac{-n\eps^2}{2CK + 2\eps e_n /3} \lt( 1 - \frac{ 2CK (\log M +\log K) + (2/3) \eps e_n (\log M + \log K)}{n\eps^2}\rt) \rt)=o(1).
}The convergence result follows from $K = o(M)$ by Lemma \ref{lem:orders} (i), $(K \log M)/n =o(1)$ by Lemma \ref{lem:orders} (ii), and $e_n \log M / n =o(1)$ by Lemma \ref{lem:orders} (iii).

Finally, we show that $A_{22}=o(1).$
\eqs{
  A_{22}
  & = P\lt( \max_{1 \le k \le K} \max_{1 \le m \le M}  \ave  \lt\vert v_{i(m,k)} \rt\vert 1\lt(\lt\vert v_{i(m,k)} \rt\vert  > e_n \rt)  \ge \eps \rt) \\
  & \le P\lt( \max_{1 \le k \le K} \max_{1 \le m \le M}  \max_{1 \le i \le n}  \lt\vert v_{i(m,k)} \rt\vert 1\lt(\lt\vert v_{i(m,k)} \rt\vert  > e_n \rt)  \ge \eps \rt) \\
  & \le P\lt( \max_{1 \le k \le K} \max_{1 \le m \le M}  \max_{1 \le i \le n}  1\lt(\lt\vert v_{i(m,k)} \rt\vert  > e_n \rt)  \rt) \\
  & \le \sum_{k=1}^K  \sum_{m=1}^{M} \sumi P \lt( \lt\vert v_{i(m,k)} \rt\vert  > e_n \rt)   \\
  & \le \frac{1}{e_n^4 } \sum_{k=1}^K  \sum_{m=1}^{M} \sumi E \lt( \lt\vert v_{i(m,k)} \rt\vert^4  \rt) = \frac{O(1)}{K}=o(1)\changed{.}
} 
\end{proof}

\subsection*{Proof of Theorem \ref{thm:uniform}}

The proof is similar to \citet{lu2015jackknife} except the last part that shows the convergence of the maximal inequality bound. Let $\dt_n := L\sqrt{n^{-1}K \log n}$ for some large constant $L<\infty$. Let $\bar{Q}_{(m,k)} \lt(\Tht_{(m,k)}\rt) := E\lt[\rho_{\tau} \lt(y_i-x_{i(m,k)}'\Tht_{(m,k)}\rt)\rt]$. We also define
\eqs{
  D(\dt_n) & := \inf_{1 \le m \le M} \inf_{\lt\Vert \Tht_{(m,k)}-\Tht^*_{(m,k)}\rt\Vert > \dt_n} \lt[ \bar{Q}_{(m,k)}\lt(\Tht_{(m,k)}\rt) - \bar{Q}_{(m,k)}\lt(\Tht^*_{(m,k)}\rt)\rt], \\
  \mathcal{S}\mk (\dt_n) & := \lt\{ \Tht\mk: \lt\Vert \Tht\mk - \Tht^*\mk \rt\Vert > \dt_n,  \lt\Vert \Tht\mk - \Tht^*\mk \rt\Vert =o(1) \rt\}.
}The same arguments in \citet{lu2015jackknife} imply that, for any $\Tht\mk \in \mathcal{S}\mk\lt(\dt_n\rt)$,
\eqs{
  &\hskip-5mm \bar{Q}\mk \lt(\Tht\mk\rt) - \bar{Q}\mk\lt(\Tht^*\mk\rt) \\
   & = E\lt[ \rho_{\tau}\lt(y_i - x'\imk \Tht\mk \rt) - \rho_{\tau}\lt(y_i - x'\imk \Tht^*\mk\rt)\rt] \\
   & = E\lt[ \rho_{\tau}\lt(\eps_i + u\imk - x'\imk \lt[\Tht\mk - \Tht^*\mk\rt]\rt) - \rho_{\tau}\lt(\eps_i - u\imk \rt)\rt] \\
   & = E \lt\{ \int_0^{x'\imk\lt[\Tht\mk-\Tht^*\mk\rt]} \lt[  1\lt\{ \eps_i + u\imk \le s \rt\} - 1\lt\{ \eps_i + u\imk \le 0 \rt\}  \rt]  ds \rt\} \\
   & = E \lt\{ \int_0^{x'\imk\lt[\Tht\mk-\Tht^*\mk\rt]} \lt[  F\lt(-u\imk+s|x_i\rt)  - F\lt(-u\imk|x_i\rt)   \rt]  ds \rt\} \\
   & \approx \frac{1}{2} \lt[\Tht\mk-\Tht^*\mk \rt]'A\mk \lt[\Tht\mk-\Tht^*\mk \rt] \ge \frac{\underline{c}_A\dt^2_n}{2}.
}The claim in (i) is established by showing that the following maximal inequality converges to zero:
\eqs{
  &\hskip-5mm P\lt(\maxi \maxk \maxm \lt\Vert \what{\Tht}\imk - \Tht^*\mk \rt\Vert \ge \dt_n \rt) =o_p(1)\changed{.}
}
We first derive the upper bound of it:
\eqs{
  &\hskip-5mm P\lt(\maxi \maxk \maxm \lt\Vert \what{\Tht}\imk - \Tht^*\mk \rt\Vert \ge \dt_n \rt) \\
  & \le nKM \maxi \maxk \maxm P\lt( \lt\Vert \what{\Tht}\imk - \Tht^*\mk \rt\Vert \ge \dt_n \rt) \\
  & \le nKM \maxi \maxk \maxm P\lt(  \bar{Q}\mk \lt(\Tht\mk\rt) - \bar{Q}\mk\lt(\Tht^*\mk\rt) \ge D(\dt_n)\rt) \\
  & \approx nKM \maxi \maxk \maxm P\lt( \mathbb{W}\imk \ge 2nD(\dt_n)\rt),
}where $\mathbb{W}\imk:=n \lt[\Tht\mk-\Tht^*\mk \rt]'A\mk \lt[\Tht\mk-\Tht^*\mk \rt]$.
We apply similar arguments in the proof of Theorem 3.2 of \citet{lu2015jackknife} to show that
\eqs{
  \mathbb{W}\imk \le \lt(\overline{c}_A\overline{c}_B/\underline{c}_A^2\rt) \lt\Vert \wtd{\bt}\imk\rt\Vert^2,
}
where $\wtd{\bt}\imk := \sqrt{n}\lt[C\mk C\mk'\rt]^{-1/2}C\mk V\mk^{-1/2}\lt[\what{\Tht}\imk - \Tht^*\mk\rt] \darrow N\lt(0,I_{l\mk}\rt)$. Let $c_{AB} := \overline{c}_A\overline{c}_B/ \underline{c}^2_A$ and $\bar{l}:=\maxk \maxm l_{(m,k)}$. Then, the above inequality for $\mathbb{W}\imk$ and the corrected version of Lemma 2.1 of \citet{shibata1981optimal,shibata1982amendments} imply that
\eqs{
  nKM  & \maxi \maxk \maxm P\lt( \mathbb{W}_{i(m,k)} \ge 2nD(\dt_n) \rt) \\
  & \le nKM \maxi \maxk \maxm P\lt( \lt\Vert \wtd{\bt}_{i(m,k)} \rt\Vert^2 \ge 2nD(\dt_n) / c_{AB} \rt) \\
  & \le \limsup_{n \rarrow \infty } nKM \maxk \maxm P \lt(\chi^2(l_{(m,k)}) \ge 2nD(\dt_n)/c_{AB} \rt) \\
 & \le \limsup_{n \rarrow \infty } nKM P \lt(\chi^2(\bar{l}) \ge 2nD(\dt_n)/c_{AB} \rt) \\
 & \le \limsup_{n \rarrow \infty } nKM P \lt(\chi^2(\bar{l}) \ge \bar{l} + \lt(n \dt_n^2\underline{c}_A/c_{AB} -\bar{l}\rt) \rt) \\
 & \le \limsup_{n \rarrow \infty } nKM \exp\lt( -0.5\lt( n \dt_n^2\underline{c}_A/c_{AB} -\bar{l} \rt) \lt(1-\log(n \dt_n^2\underline{c}_A/(\bar{l}c_{AB}) )/(n \dt_n^2\underline{c}_A/(\bar{l}c_{AB}) -1) \rt) \rt) \\
 & = o(1)\changed{.}
}For the last equality, note first that $\log(n \dt_n^2\underline{c}_A/(\bar{l}c_{AB}) )/(n \dt_n^2\underline{c}_A/(\bar{l}c_{AB}) -1)  =o(1)$  by Assumption \ref{a-3}(ii). The leading term  becomes
\eqs{
  nKM \exp\lt( -0.5\lt( n \dt_n^2\underline{c}_A/c_{AB}  \rt) \rt)
  & = nKM n^{ -0.5\lt( L^2 K \underline{c}_A^3/\lt(\overline{c}_A \overline{c}_B\rt)\rt) } \\
  & \ll KK! n^{ 1 -0.5\lt( L^2 K \underline{c}_A^3/\lt(\overline{c}_A \overline{c}_B\rt)\rt) } \\
  & \ll KK^K n^{ 1 -0.5\lt( L^2 K \underline{c}_A^3/\lt(\overline{c}_A \overline{c}_B\rt)\rt) } \\
  & = K^{K+1} n^{ 1 -0.5\lt( L^2 K \underline{c}_A^3/\lt(\overline{c}_A \overline{c}_B\rt)\rt) } \\
  & \le C (\log n)^{K+1} n^{ 1 -0.5\lt( L^2 K \underline{c}_A^3/\lt(\overline{c}_A \overline{c}_B\rt)\rt) } \\
  & =o(1)
}where $C<\infty$ is a generic constant. The second line holds by the definition of $M$, the third line holds by the fact that $K! \ll K^K$, the fifth line holds by Assumption \ref{a-3}(ii), and the last convergence result holds by \ref{a-3}(ii) and by taking some large $L$.

Therefore, we establish the result in (i). Analogously, we can prove the result in (ii). \qed


\subsection*{Proof of Theorem \ref{thm:subsample}}
Using the definition of $\what{y}(k)$ and $\wtd{y}(k)$ and the triangular inequality, we have
\eqs{
	\max_{1 \le k \le K} &  \lt\vert \what{y}(k) - \wtd{y}(k)\rt\vert \\
	& = \max_{1 \le k \le K} \lt\vert \frac{1}{M} \sum_{m=1}^M x'_{(m,k)} \what{\Theta}_{(m,k)}
		- \frac{1}{M_{max}} \sum_{m' \in \mathcal{M}_{max}} x_{(m',k)}' \what{\Theta}_{(m',k)}\rt\vert	\\
	& = \max_{1 \le k \le K} \lt\vert \frac{1}{M} \sum_{m=1}^M \lt( x'_{(m,k)} \what{\Theta}_{(m,k)} - y^*\rt)
		- \frac{1}{M_{max}} \sum_{m' \in \mathcal{M}_{max}} \lt( x_{(m',k)}' \what{\Theta}_{(m',k)} - y^*\rt)\rt\vert \\
	& \le \max_{1 \le k \le K}   \lt\vert \frac{1}{M} \sum_{m=1}^M \lt( x'_{(m,k)} \what{\Theta}_{(m,k)} - y^*\rt) \rt\vert 
		+ \max_{1 \le k \le K} \lt\vert \frac{1}{M_{max}} \sum_{m' \in \mathcal{M}_{max}} \lt( x_{(m',k)}' \what{\Theta}_{(m',k)} - y^*\rt)\rt\vert  \\
	& \equiv \max_{1 \le k \le K}  EQ_1 + \max_{1 \le k \le K}  EQ_2.
}
We first investigate $EQ_1$:
\eqs{
	\max_{1 \le k \le K} EQ_1 & \le \max_{1 \le k \le K}  \lt\vert \frac{1}{M} \sum_{m=1}^M \lt( x'_{(m,k)} {\Theta}_{(m,k)}^* - y^*_k\rt) \rt\vert
			+ \max_{1 \le k \le K}  \lt\vert \frac{1}{M} \sum_{m=1}^M x'_{(m,k)} \lt( \what{\Theta}_{(m,k)} - \Theta_{(m,k)}^*\rt) \rt\vert \\
		 &  \le \max_{1 \le k \le K}  \lt\vert \frac{1}{M} \sum_{m=1}^M \lt( x'_{(m,k)} {\Theta}_{(m,k)}^* - y^*_k\rt) \rt\vert
			+ \max_{1 \le k \le K}  \frac{1}{M} \sum_{m=1}^M \lt\Vert x_{(m,k)} \rt\Vert \lt\Vert \what{\Theta}_{(m,k)} - \Theta_{(m,k)}^*\rt\Vert  \\
		 &  \le \max_{1 \le k \le K}  \lt\vert \frac{1}{M} \sum_{m=1}^M \lt( x'_{(m,k)} {\Theta}_{(m,k)}^* - y^*_k\rt)  \rt\vert \\
		 & \hskip30pt
			+ \lt(\max_{1 \le k \le K}  \max_{1 \le m \le M} \lt\Vert x_{(m,k)} \rt\Vert \rt) \lt( \max_{1 \le k \le K}  \max_{ 1 \le m \le M} \lt\Vert \what{\Theta}_{(m,k)} - \Theta_{(m,k)}^*\rt\Vert \rt) \\
		 & = o_p(1) + O_p(1) \cdot o_p(1) \\
		 & = o_p(1)\changed{.}
}The first inequality holds from the triangular inequality and the second one from the Cauchy-Schwartz inequality. The final inequality holds from the uniform convergence and boundedness assumptions, and Theorem \ref{thm:uniform} (ii) above. Similarly, we can show $EQ_2=o_p(1)$. \qed


\subsection*{Proof of Theorem \ref{thm:opt}}

It suffices to show that, with $\mathcal{K}:=\left\{ 1,...,K_{n}\right\}$,
\eq{
\sup_{k \in \mathcal{K}} \lt\vert \frac{CV_n(k) - FPE_n(k)}{FPE_n(k)}\rt\vert =o_p(1)\changed{.} \label{eq:opt-main}
}

We first expand the numerator by applying Knight's identity repeatedly.
\eqs{
& CV_n(k)-FPE_n(k)  \\
= & \lt\{ \frac{1}{n}\sum_{i=1}^n \lt[ \rho_{\tau} \left(y_i -  M^{-1}\sum_{m=1}^{M} x_{i(m,k)}' \widehat {\Theta}_{i(m,k)}\right)
- \rho_{\tau}(\eps_i) \rt]  \rt\} \\
& -  \lt\{ FPE_n(k) - E\lt[\rho_{\tau}(\eps) \rt] \rt\}
+\ave \lt\{  \rho_{\tau}(\eps_i) - E\lt[\rho_{\tau}(\eps) \rt] \rt\} \\
= & \ave \lt[ \mu_i - M^{-1}\sum_{m=1}^{M} x_{i(m,k)}' \what{\Tht}_{i(m,k)}\rt] \psi_{\tau}(\eps_i) \\
& +  \ave \int_0^{M^{-1} \summ x_{i(m,k)}' \what{\Tht}_{i(m,k)} - \mu_i } \lt[1\lt\{\eps_i \le s\rt\} - 1\lt\{\eps_i \le 0 \rt\} \rt] ds\\
& -  E \lt[ \int_0^{M^{-1} \summ x_{(m,k)}'\what{\Tht}_{(m,k)} -  \mu } \lt[1\lt\{\eps \le s\rt\} - 1\lt\{\eps \le 0 \rt\} \rt] ds \vert \mathcal{D}_n \rt] \\
& +\ave \lt\{  \rho_{\tau}(\eps_i) - E\lt[\rho_{\tau}(\eps) \rt] \rt\} \\
= & \ave \lt[ \mu_i - M^{-1}\sum_{m=1}^{M} x_{i(m,k)}' \what{\Tht}_{i(m,k)}\rt] \psi_{\tau}(\eps_i) \\
& +  \ave \int_0^{M^{-1} \summ x_{i(m,k)}' \what{\Tht}_{i(m,k)} - \mu_i } \lt[1\lt\{\eps_i \le s\rt\} - 1\lt\{\eps_i \le 0 \rt\} - F(s|x_i) +F(0|x_i) \rt] ds\\
& +  \ave \Big\{ \int_0^{M^{-1} \summ x_{i(m,k)}' \what{\Tht}_{i(m,k)} - \mu_i } \lt[F(s|x_i) - F(0|x_i) \rt] ds \\
& \hskip50pt -E_{x_i} \lt[ \int_0^{M^{-1}\summ x_{i(m,k)}' \what{\Tht}_{i(m,k)} - \mu_i } \lt[F(s|x_i) - F(0|x_i) \rt] ds \rt] \Big\} \\
& + \ave \Bigg\{ E_{x} \lt[ \int_0^{M^{-1} \summ x_{(m,k)}' \what{\Tht}_{i(m,k)} - \mu } \lt[F(s|x) - F(0|x) \rt] ds \rt]  \\
& \hskip50pt - E_{x} \lt[ \int_0^{M^{-1}\summ x_{(m,k)}'\what{\Tht}_{(m,k)} -  \mu } \lt[F(s|x) - F(0|x) \rt] ds \rt] \Bigg\}\\
& +\ave \lt\{  \rho_{\tau}(\eps_i) - E\lt[\rho_{\tau}(\eps) \rt] \rt\} \\
& \equiv CV_{1n} + CV_{2n} + CV_{3n} + CV_{4n} + CV_{5n}\changed{.}
}
\normalsize
It is straightforward to derive all terms except $CV_{4n}$. We need the following two results to get $CV_{4n}$. Let $E_x$ be an expectation with respect to a random variable $x$.
\eq{
  \begin{split}
  & \ave E_{x_i} \lt[ \int_0^{M^{-1}\sum_{m=1} x_{i(m,k)}' \hat{\Tht}_{i(m,k)} - \mu_i} [F(s|x_i)-F(0|x_i)]ds \rt] \\
  &\hskip15pt = \ave E_{x} \lt[ \int_0^{M^{-1}\sum_{m=1}^M w_m x_{(m,k)}' \hat{\Tht}_{i(m,k)} - \mu} [F(s|x)-F(0|x)]ds \rt], \label{eq:iden-1}
  \end{split}
  \\
  \begin{split}
  & E\lt[ \int_0^{M^{-1}\sum_{m=1}^M x_{(m,k)}' \hat{\Tht}_{(m,k)} - \mu} [1\{\eps \le s\} - 1\{\eps \le 0 \}]ds | D_n \rt] \\
  &\hskip15pt = E_x\lt[\int_0^{M^{-1}\sum_{m=1}^M  x_{(m,k)}' \hat{\Tht}_{(m.k)} - \mu}  [F(s|x) - F(0|x)] ds  \rt] \label{eq:iden-2}.
  \end{split}
}
The identity \eqref{eq:iden-1} follows from the fact that $\hat{\Tht}_{i(m)}$ does not depend on the $i$-th observation. The second result \eqref{eq:iden-2} comes from
\eqs{
  & E\lt[ \int_0^{M^{-1} \sum_{m=1}^M  x_{(m,k)}' \hat{\Tht}_{(m,k)} - \mu} [1\{\eps \le s\} - 1\{\eps \le 0 \}]ds | D_n \rt] \\
  & = \int_{(x,\eps)} \int_0^{M^{-1}\sum_{m=1}^M  x_{(m,k)}' \hat{\Tht}_{(m,k)} - \mu} [1\{\eps \le s\} - 1\{\eps \le 0 \}]ds f(x, \eps |D_n) dx d\eps \\
  & = \int_{(x,\eps)} \int_0^{M^{-1} \sum_{m=1}^M x_{(m,k)}' \hat{\Tht}_{(m,k)} - \mu} [1\{\eps \le s\} - 1\{\eps \le 0 \}]ds f(x, \eps) dx d\eps \\
  & = \int_{x} \int_0^{M^{-1} \sum_{m=1}^M  x_{(m,k)}' \hat{\Tht}_{(m,k)} - \mu}  \int_{\eps} [1\{\eps \le s\} - 1\{\eps \le 0 \}] f(\eps|x)f(x)  d\eps dxds\\
  & = \int_{x} \int_0^{ M^{-1} \sum_{m=1}^M x_{(m,k)}' \hat{\Tht}_{(m,k)} - \mu}  [F(s|x) - F(0|x)] f(x) dx ds \\
  & = E_x\lt[ \int_0^{M^{-1} \sum_{m=1}^M  x_{(m,k)}' \hat{\Tht}_{(m,k)} - \mu}  [F(s|x) - F(0|x)] ds \rt]\changed{.}
}The second equality holds by the independence of the sample $\{x_i,\eps_i\}$ and the generic random variable $(x,\eps)$.

We are now ready to prove \eqref{eq:opt-main}. We first show that the denominator of \eqref{eq:opt-main} is uniformly bounded above zero and show the uniform convergence of $CV_{1n}, \ldots, CV_{5n}$.

\textbf{Claim 1: $\min_{k \in K} FPE_n(k) \ge E[\rho_{\tau}(\eps)] - o_p(1)$.}  This results shows that the denominator of the LHS in \eqref{eq:opt-main} is bounded above zero.
Let $u_k :=\mu - M^{-1} \summ  x_{(m,k)}'\Tht_{(m,k)}^*$.
\eqs{
&FPE_n(k) - E[\rho_{\tau}(\eps+u_k)] \\
& = E\lt[ \rhot \lt( \eps + u_k - \avem x_{(m,k)}' \lt(\what{\Tht}_{(m,k)} - \Tht^*_{(m,k)} \rt)\rt) - \rhot\lt( \eps + u_k \rt) \vert \mathcal{D}_n \rt] \\
& = E\lt[ \int_0^{\avem x_{(m,k)}'\lt(\what{\Tht}_{(m,k)} - \Tht^*_{(m,k)} \rt)} \lt[1\lt\{\eps+u_k \le s \rt\} - 1\lt\{ \eps + u_k \le 0 \rt\}\rt] ds \vert \mathcal{D}_n \rt] \\
& = E_x \lt[ \int_0^{\avem x_{(m,k)}'\lt(\what{\Tht}_{(m,k)} - \Tht^*_{(m,k)} \rt)} \lt[ F(s-u_k |x)- F(-u_k|x)\rt] ds \rt] \\
& = E_x \lt[   \int_0^{\avem x_{(m,k)}'\lt(\what{\Tht}_{(m,k)} - \Tht^*_{(m,k)} \rt)} f(-u_k|x) s ds
\rt] +o_p(1)\\
& = 2^{-1}E_x \lt[   f(-u_k|x) \lt[\avem x_{(m,k)}'\lt(\what{\Tht}_{(m,k)} - \Tht^*_{(m,k)} \rt) \rt]^2  \rt] +o_p(1) \\
& \le 2^{-1}E_x \lt[   f(-u_k|x) \avem \lt[ x_{(m,k)}'\lt(\what{\Tht}_{(m,k)} - \Tht^*_{(m,k)} \rt) \rt]^2  \rt] +o_p(1)  \\
&= 2^{-1} \lt\{   \avem  \lt(\what{\Tht}_{(m,k)} - \Tht^*_{(m,k)} \rt)' E_x \lt[f(-u_k|x)  x_{(m,k)} x_{(m,k)}'  \rt] \lt(\what{\Tht}_{(m,k)} - \Tht^*_{(m,k)} \rt)  \rt\} +o_p(1)  \\
& \le \frac{\bar{c}_A}{2}\max_{1\le k \le K} \max_{1\le m \le M} \Vert \what{\Tht}_{(m,k)} - \Tht^*_{(m,k)} \Vert^2 + o_p(1) = o_p(1).
}

\textbf{Claim 2: $\sup_{k \in \mathcal{K}} | CV_{1n} (k)| =o_p(1)$.}
\eqs{
  CV_{1n}(k) &= \ave \lt[ \mu_i - M^{-1}\sum_{m=1}^{M} x_{i(m,k)}' {\Tht}_{i(m,k)}^* \rt] \psi_{\tau}(\eps_i) \\
  & \hskip30pt -\ave \lt[  M^{-1}\sum_{m=1}^{M} x_{i(m,k)}' \lt( \what{\Tht}_{i(m,k)} - {\Tht}_{i(m,k)}^* \rt) \rt] \psi_{\tau}(\eps_i) \\
  & \equiv CV_{1n,1} + CV_{1n,2}\changed{.}
}
We first show that $\sup_{k \in \mathcal{K}} CV_{1n,1} =o_p(1)$. Let $b_{i(m,k)}= \mu_i - x_{i(m,k)}'\Tht_{i(m,k)}^*$ and $e_n=(MnK^2)^{1/4}$. Note that
\eqs{
  P\lt( \max_{1\le k\le K} \lt\vert CV_{1n,1}  \rt\vert \ge 2\eps \rt)
  & \le P \lt( \max_{1\le k\le K}  \frac{1}{nM} \sumi \summ \lt\vert b_{i(m,k)} \rt\vert \ge 2\eps \rt) \\
  & \le P\lt( \max_{1\le k\le K} \frac{1}{nM} \sumi \summ \lt\vert b_{i(m,k)} \rt\vert 1\lt( \lt\vert b_{i(m,k)} \rt\vert \le e_n \rt) \ge \eps \rt) \\
  & \hskip15pt + P\lt( \max_{1\le k\le K} \frac{1}{nM} \sumi \summ \lt\vert \mu_i - x_{i(m,k)}'\Tht_{i(m,k)}^* \rt\vert 1\lt(\lt\vert b_{i(m,k)} \rt\vert > e_n\rt) \ge \eps \rt) \\
  & \equiv CV_{1n,11} + CV_{1n,12}.
}We next show that $CV_{1n,11}=o(1)$ and $CV_{1n,12}=o(1)$, respectively.
\eqs{
  CV_{1n,11}
  & \le K \max_{1 \le k \le K} P\lt( \frac{1}{nM} \sumi \summ \lt\vert b_{i(m,k)} \rt\vert 1\lt( \lt\vert b_{i(m,k)} \rt\vert \le e_n \rt) \ge \eps \rt) \\
  & \le 2 K \exp\lt(- \frac{nM \eps^2}{ 2K C + 2\eps e_n /3}\rt) \\
  & \le 2 \exp \lt(- \frac{nM \eps^2}{ 2K C + 2\eps e_n /3} + \log K\rt) \\
  & = 2 \exp \lt(- \frac{nM \eps^2}{ 2K C + 2\eps e_n /3} \lt(1 - \frac{ (2K C + 2\eps e_n /3)\log K} {nM \eps^2} \rt) \rt) =o(1).
}The last convergence result follows from the order conditions in Assumption \ref{a-3}.
\eqs{
  CV_{1n,12}
  & \le P\lt( \max_{1 \le k \le K} \max_{1 \le i \le n} \max_{1 \le m \le M}  \lt\vert b_{i(m,k)}\rt\vert >e_n   \rt) \\
  & \le \sumk \sumi \summ P\lt( \lt\vert b_{i(m,k)}\rt\vert >e_n \rt)\\
  & \le \frac{1}{e_n^4} \sumk \sumi \summ E\lt[ \lt\vert b_{i(m,k)} \rt\vert^4  1\lt( \lt\vert b_{i(m,k)}\rt\vert^4 > e_n^4 \rt)  \rt] =o(1)\changed{.}
}

We next turn our attention to $CV_{1n,2}$:
\eqs{
  \sup_{k \in \mathcal{K}} \lt\vert CV_{1n,2} \rt\vert
  & \le \sup_{k \in \mathcal{K}} \frac{1}{n M} \sumi   \sum_{m=1}^{M} \lt\vert x_{i(m,k)}' \lt( \what{\Tht}_{i(m,k)} - {\Tht}_{i(m,k)}^* \rt)  \psi_{\tau}(\eps_i)\rt\vert \\
  & \le \sup_{k \in \mathcal{K}} \frac{1}{n M} \sumi   \summ  \lt\vert x_{i(m,k)}' \lt( \what{\Tht}_{i(m,k)} - {\Tht}_{i(m,k)}^* \rt)  \rt\vert \\
  & \le \sup_{k \in \mathcal{K}} \frac{1}{n M} \sumi   \summ  \lt\Vert x_{i(m,k)} \rt\Vert  \lt\Vert \what{\Tht}_{i(m,k)} - {\Tht}_{i(m,k)}^*  \rt\Vert \\
  & = \sup_{k \in \mathcal{K}} \frac{1}{M}    \summ \lt\{ \ave \lt\Vert x_{i(m,k)} \rt\Vert  \lt\Vert \what{\Tht}_{i(m,k)} - {\Tht}_{i(m,k)}^*  \rt\Vert \rt\} \\
  & \le \sup_{k \in \mathcal{K}} \frac{1}{M}    \summ \lt\{ \ave \lt\Vert x_{i(m,k)} \rt\Vert  \lt(\max_{1\le i \le n}\lt\Vert \what{\Tht}_{i(m,k)} - {\Tht}_{i(m,k)}^*  \rt\Vert \rt) \rt\} \\
  & = \sup_{k \in \mathcal{K}} \frac{1}{M}    \summ \lt\{ \lt(\max_{1\le i \le n}\lt\Vert \what{\Tht}_{i(m,k)} - {\Tht}_{i(m,k)}^*  \rt\Vert \rt)  \ave \lt\Vert x_{i(m,k)} \rt\Vert  \rt\} \\
  & \le \sup_{k \in \mathcal{K}} \frac{1}{M}    \summ  \max_{1\le m \le M}\lt\{ \lt(\max_{1\le i \le n}\lt\Vert \what{\Tht}_{i(m,k)} - {\Tht}_{i(m,k)}^*  \rt\Vert \rt)  \ave \lt\Vert x_{i(m,k)} \rt\Vert  \rt\} \\
  & = \sup_{k \in \mathcal{K}}  \max_{1\le m \le M}\lt\{ \lt(\max_{1\le i \le n}\lt\Vert \what{\Tht}_{i(m,k)} - {\Tht}_{i(m,k)}^*  \rt\Vert \rt)  \ave \lt\Vert x_{i(m,k)} \rt\Vert  \rt\} \frac{1}{M}    \summ 1\\
  & =\sup_{k \in \mathcal{K}}  \max_{1\le m \le M}\lt\{ \lt(\max_{1\le i \le n}\lt\Vert \what{\Tht}_{i(m,k)} - {\Tht}_{i(m,k)}^*  \rt\Vert \rt)  \ave \lt\Vert x_{i(m,k)} \rt\Vert  \rt\} \\
  & = \lt\{\sup_{k \in \mathcal{K}} \max_{1\le i \le n} \max_{1\le m \le M} \lt\Vert \what{\Tht}_{i(m,k)} - {\Tht}_{i(m,k)}^*  \rt\Vert \rt\}  \lt\{ \sup_{k \in \mathcal{K}} \max_{1\le m \le M} \ave \lt\Vert x_{i(m,k)} \rt\Vert  \rt\} \\
  & = O_p\lt(\sqrt{n^{-1}K \log n}\rt) O_p\lt(\sqrt{K}\rt)= o_p(1).
} The convergence results follow from Theorem \ref{thm:uniform}, Lemma \ref{lem:x-rate}, and Assumption \ref{a-3}\changed{.}

\textbf{Claim 3: $\sup_{k \in \mathcal{K}} | CV_{2n} (k)| =o_p(1)$.}
\eqs{
  \lt\vert CV_{2n} (k)\rt\vert \le \lt\vert CV_{2n,1} (k)\rt\vert +  \lt\vert CV_{2n,2} (k)\rt\vert
}where
\eqs{
  CV_{2n,1} (k)& =   \ave  \int_0^{M^{-1} \summ x_{i(m,k)}' {\Tht}_{(m,k)}^{*} - \mu_i } \lt[1\lt\{\eps_i \le s\rt\} - 1\lt\{\eps_i \le 0 \rt\} - F(s|x_i) +F(0|x_i) \rt] ds \\
\mbox{and}\\
  CV_{2n,2} (k)& =   \ave \int_{M^{-1} \summ x_{i(m,k)}' {\Tht}_{(m,k)}^{*} - \mu_i }^{M^{-1} \summ x_{i(m,k)}' \what{\Tht}_{i(m,k)} - \mu_i } \lt[1\lt\{\eps_i \le s\rt\} - 1\lt\{\eps_i \le 0 \rt\} - F(s|x_i) +F(0|x_i) \rt] ds\changed{.}
}
Since $1\lt\{\eps_i \le s\rt\} - 1\lt\{\eps_i \le 0 \rt\} - F(s|x_i) +F(0|x_i) \le 2$, we have
\eqs{
  \lt\vert CV_{2n,1} (k)\rt\vert \le\frac{2}{nM} \sumi \summ \lt\vert x_{i(m,k)}' {\Tht}_{(m,k)}^{*} - \mu_i \rt\vert.
}Thus, $\sup_{k \in \mathcal{K}} \lt\vert CV_{2n,1} (k) \rt\vert = o_p(1)$ follows from the same arguments used for $CV_{1n,1}$ above.

We next investigate $CV_{2n,2}$. We have
\eqs{
  \sup_{k \in \mathcal{K}} \lt\vert CV_{2n,2} (k) \rt\vert &\le \sup_{k \in \mathcal{K}} \frac{2}{n M} \sumi \summ \lt\vert x_{i,(m,k)}' (\what{\Tht}_{i(m,k)} - \Tht_{(m,k)}^{*}) \rt\vert \\
  & \le \sup_{k \in \mathcal{K}} \frac{2}{n M} \sumi \summ  \lt\Vert x_{i,(m,k)} \rt\Vert  \lt\Vert \what{\Tht}_{i(m,k)} - \Tht_{(m,k)}^{*} \rt\Vert \\
  & \le 2 \sup_{k \in \mathcal{K}} \max_{1\le i \le n} \max_{1 \le m \le M} \lt\Vert \what{\Tht}_{i(m,k)} - \Tht_{(m,k)}^{*} \rt\Vert \sup_{k \in \mathcal{K}}  \max_{1 \le m \le M} \ave \lt\Vert x_{i,(m,k)} \rt\Vert \\
  & = O_p\lt( \sqrt{n^{-1}K \log n} \rt) \cdot O_p\lt( \sqrt{K} \rt) =o_p(1).
}

\textbf{Claim 4: $\sup_{k \in \mathcal{K}} | CV_{3n} (k)| =o_p(1)$.}
\eqs{
  \lt\vert CV_{3n} (k)\rt\vert \le \lt\vert CV_{3n,1} (k)\rt\vert +  \lt\vert CV_{3n,2} (k)\rt\vert
}where
\eqs{
  CV_{3n,1} (k)& =   \ave \Bigg\{ \int_0^{M^{-1} \summ x_{i(m,k)}' {\Tht}_{(m,k)}^{*} - \mu_i } \lt[F(s|x_i) - F(0|x_i) \rt] ds \\
& \hskip30pt -E_{x_i} \lt[ \int_0^{M^{-1}\summ x_{i(m,k)}' {\Tht}_{(m,k)}^{*} - \mu_i } \lt[F(s|x_i) - F(0|x_i) \rt] ds \rt] \Bigg\} \\
\mbox{and}\\
  CV_{3n,2} (k)& =   \ave \Bigg\{ \int_{M^{-1} \summ x_{i(m,k)}' {\Tht}_{(m,k)}^{*} - \mu_i }^{M^{-1} \summ x_{i(m,k)}' \what{\Tht}_{i(m,k)} - \mu_i } \lt[F(s|x_i) - F(0|x_i) \rt] ds \\
& \hskip30pt -E_{x_i} \lt[ \int_{M^{-1}\summ x_{i(m,k)}' {\Tht}_{(m,k)}^{*} - \mu_i }^{M^{-1} \summ x_{i(m,k)}' \what{\Tht}_{i(m,k)} - \mu_i } \lt[F(s|x_i) - F(0|x_i) \rt] ds \rt] \Bigg\}\changed{.}
}The proof of $\sup_{k \in \mathcal{K} }|CV_{3n,1}|=o_p(1)$ is similar to that of $\sup_{k \in \mathcal{K}}|CV_{1n,1}|=o_p(1)$ in Claim 2 and is omitted. Note that
\eqs{
  \lt\vert CV_{3n,2} (k) \rt\vert
  & \le \ave \lt\vert M^{-1} \summ x_{i(m,k)}'\lt( \what{\Tht}_{i(m,k)} - \Tht_{(m,k)}^* \rt) \rt\vert \\
  & \hskip30pt \ave E_{x_i} \lt\vert M^{-1} \summ x_{i(m,k)}'\lt( \what{\Tht}_{i(m,k)} - \Tht_{(m,k)}^* \rt) \rt\vert \\
  & \equiv CV_{3n,21}(k) + CV_{3n,22}(k)\changed{.}
}The proof of $\sup_{k \in \mathcal{K} }|CV_{3n,21}|=o_p(1)$ is similar to that of $\sup_{k \in \mathcal{K}}|CV_{2n,2}|=o_p(1)$ in Claim 3 and is also omitted. It remains to show that $\sup_{k \in \mathcal{K} }|CV_{3n,22}|=o_p(1)$. Note that
\eqs{
  \sup_{k \in \mathcal{K}} CV_{3n,22}(k)
  & \le \sup_{k \in \mathcal{K}} \frac{1}{n M} \sumi \summ E_{x_i} \lt\vert x_{i(m,k)}' \lt( \what{\Tht}_{i(m,k)} - \Tht_{(m,k)}^*  \rt) \rt\vert \\
  & = \sup_{k \in \mathcal{K}} \frac{1}{n M} \sumi \summ E_{x_i}  \lt[\lt( \what{\Tht}_{i(m,k)} - \Tht_{(m,k)}^* \rt)'  x_{i(m,k)} x_{i(m,k)}' \lt( \what{\Tht}_{i(m,k)} - \Tht_{(m,k)}^*  \rt) \rt]^{1/2}  \\
  & \le \sup_{k \in \mathcal{K}} \frac{1}{n M} \sumi \summ   \lt[\lt( \what{\Tht}_{i(m,k)} - \Tht_{(m,k)}^* \rt)' E_{x_i} \lt[  x_{i(m,k)} x_{i(m,k)}' \rt] \lt( \what{\Tht}_{i(m,k)} - \Tht_{(m,k)}^*  \rt) \rt]^{1/2}  \\
  & \le \sup_{k \in \mathcal{K}} \frac{1}{n M} \sumi \summ   \lt[ \ld_{max} \lt( E_{x_i} \lt[  x_{i(m,k)} x_{i(m,k)}' \rt] \rt) \rt]^{1/2}  \lt\Vert \what{\Tht}_{i(m,k)} -  \Tht_{(m,k)}^*  \rt\Vert \\
  & \le \lt(\max_{k \in \mathcal{K}} \max_{1\le m \le M} \lt[ \ld_{max} \lt( E_{x_i} \lt[  x_{i(m,k)} x_{i(m,k)}' \rt] \rt) \rt]^{1/2} \rt) \\
  & \hskip30pt \times
  \lt(\max_{1\le i \le n} \max_{k \in \mathcal{K}} \max_{1\le m \le M} \lt\Vert \what{\Tht}_{i(m,k)} -  \Tht_{(m,k)}^*  \rt\Vert \rt)\\
  & = o_p(1)
}\changed{by} the triangle inequality, $\vert x \vert = (x^2)^{1/2}$, the Jensen's inequality, $A'BA \le \ld_{max}(B)A'A$ for any real symmetric matrix $B$.

\textbf{Claim 5: $\sup_{k \in \mathcal{K}} | CV_{4n} (k)| =o_p(1)$.}
\eqs{
  \lt\vert CV_{4n} (k) \rt\vert & = \lt\vert \ave E_{x} \lt[ \int_{M^{-1}\summ x_{(m,k)}'\what{\Tht}_{(m,k)} -  \mu } ^{M^{-1} \summ x_{(m,k)}' \what{\Tht}_{i(m,k)} - \mu } \lt[F(s|x) - F(0|x) \rt] ds \rt] \rt\vert \\
  & \le  \ave E_{x} \lt\vert \lt[ \int_{M^{-1}\summ x_{(m,k)}'\what{\Tht}_{(m,k)} -  \mu } ^{M^{-1} \summ x_{(m,k)}' \what{\Tht}_{i(m,k)} - \mu } \lt[F(s|x) - F(0|x) \rt] ds \rt] \rt\vert \\
  & \le  \ave E_{x} \lt\vert   \lt({M^{-1} \summ x_{(m,k)}' \what{\Tht}_{i(m,k)} - \mu } \rt) - \lt( {M^{-1}\summ x_{(m,k)}'\what{\Tht}_{(m,k)} -  \mu } \rt) \rt\vert \\
  & = \ave E_{x} \lt\vert    M^{-1} \summ x_{(m,k)}' \lt(\what{\Tht}_{i(m,k)} - \what{\Tht}_{(m,k)} \rt)  \rt\vert \\
  & = o_p(1)
}\changed{by} the triangle inequality, $F(s|x) - F(0|x) \le 1$, and the similar arguments in the proof of $\sup_{k in \mathcal{K}} |CV_{3n,22}(k)|=o_p(1)$.

\textbf{Claim 6: $CV_{5n}=o_p(1)$.} Since $CV_{5n}$ does not depend on $k$, this result follows from the weak law of large \changed{numbers}. \qed

\subsection*{Proof of Theorem \ref{thm:prediction-efficiency-bound}}
(i) There exists $\tilde{w}$ between $\bar{w}$ and $w^*$ such that
	\eqs{
		F(\bar{w})
		& = F(w^*) + \triangledown_1 F(w^*)'(\bar{w} - w^*) + \frac{1}{2} (\bar{w} - w^*)' \triangledown_2 F(\tilde{w}) (\bar{w} - w^*) \\
		& = F(w^*) - \tilde{\ld} \cdot 1_M'(\bar{w} - w^*) + \frac{1}{2} (\bar{w} - w^*)' \triangledown_2 F(\tilde{w}) (\bar{w} - w^*) \\
		& = F(w^*) + \frac{1}{2} (\bar{w} - w^*)' \triangledown_2 F(\tilde{w}) (\bar{w} - w^*)
	}where $\tilde{\ld}$ is a Lagrange multiplier from the constraint optimization problem:
	\eq{
		w^* = \argmax_{w \in \mathbb{R}^M } F(w) + \tilde{\ld} \cdot (1_M'w - 1). \label{eq:const_opt}
	}Note that the second equality above comes from the first order condition for $w^*$ and that the third equality \changed{holds} by the normalization, $1_M'w=1$ for any weight $w$. We investigate the upper bound of the quadratic term:
	\eqs{
		\frac{1}{2} (\bar{w} - w^*)' \triangledown_2 F(\tilde{w}) (\bar{w} - w^*)
			& =    2^{-1} (\bar{w}' \triangledown_2 F(\tilde{w}) \bar{w} - 2 \bar{w}' \triangledown_2 F(\tilde{w}) w^* + w^{*\prime} \triangledown_2 F(\tilde{w})w^* ) \\
			& \equiv 2^{-1} (I + II + III)\changed{.}
	}Since $\triangledown_2 F(\tilde{w})$ is \changed{an} $M\times M$ symmetric matrix, we can factorize it as $S\Lambda S'$, where $\Lambda$ is a diagonal matrix composed of the eigenvalues $\{\ld_m\}$ and $S$ is composed of the corresponding orthonormal eigenvectors $\{s_m\}$. Note that
	\eqs{
		I 	& =   M^{-2} 1_M' \triangledown_2 F(\tilde{w})  1_M \\
			& =   M^{-2} 1_M' \lt(\sum_{m=1}^M \ld_m s_m s_m' \rt)  1_M \\
			& \le M^{-2} \bar{\ld}_{max} \Vert 1_M \Vert^2 \\
			& =   M^{-1} \bar{\ld}_{max}.\\
		II  & \le 2 \vert \bar{w}' \triangledown_2 F(\tilde{w}) w^* \vert \\
			& = 2 M^{-1} \lt\vert 1_M' \lt(\sum_{m=1}^M \ld_m s_m s_m' \rt) w^* \rt\vert \\
			& \le 2 M^{-1} \bar{\ld}_{max} \lt\vert 1_M' w^* \rt\vert \\
			& =   2 M^{-1} \bar{\ld}_{max} \\
		III & =  w^{*\prime}  \lt(\sum_{m=1}^M \ld_m s_m s_m' \rt)  w^* \\
			& \le \bar{\ld}_{max} \Vert w^* \Vert^2 \\
			& \le \bar{\ld}_{max} \Vert w^* \Vert_1^2 \\
			& = \bar{\ld}_{max}\changed{,}
	}where $\Vert \cdot \Vert_1$ denotes $\ell_1$-norm. Therefore, we have
	\eqs{
		\frac{1}{2} (\bar{w} - w^*)' \triangledown_2 F(\tilde{w}) (\bar{w} - w^*)
			& \le 2^{-1}\bar{\ld}_{max} \lt(1 + 3M^{-1}\rt),
	}which establishes the desired result.

	(ii) Using the similar arguments above, we have
	\changed{
    \eqs{
		F(\hat{w}) 	& = 	F(w^*) + 2^{-1} \hat{\eta}' \triangledown_2 F(\tilde{w}) \hat{\eta} \\
					& = 	F(w^*) + 2^{-1} \hat{\eta}' \lt(\sum_{m=1}^M \tilde{\ld}_m \tilde{s}_m \tilde{s}_m' \rt) \hat{\eta} \\
					& \le 	F(w^*) + 2^{-1} \bar{\ld}_{max} \lt\Vert \hat{\eta} \rt\Vert^2, 
	}
    where $\hat{\eta}$, $\tilde{w}$, $\tilde{\ld}_m$, and $\tilde{s}_m$ are all random objects in the second equality. The third inequality holds almost surely by the definition of $\bar{\ld}_{max}$. We establish the desired result by noting that $E\lt\Vert \hat{\eta} \rt\Vert^2 \le M \bar{\sigma}_{\eta}^2$.}\qed

\subsection*{Proof of Corollary \ref{cor:equivalence}}

	Following similar arguments in Theorem \ref{thm:prediction-efficiency-bound}, we only need to show that $w^{*\prime}  \Sigma  w^* \rightarrow 0$ as $M\rightarrow \infty$. Note that we have a closed-form solution $w^*=(1_M' \Sigma^{-1} 1_M)^{-1} \Sigma^{-1} 1_M$ for the optimization problem in \eqref{eq:const_opt}. Then, we have
	\eqs{
		w^{*\prime}  \Sigma  w^* & = (1_M' \Sigma^{-1} 1_M)^{-1}.
	}Abusing notation on eigenvalues/eigenvectors, we have
	\eqs{
		1_M' \Sigma^{-1} 1_M & =   1_M'  \lt(\sum_{m=1}^M \ld^{-1}_m s_m s_m' \rt)  1_M \\
							 & \ge \bar{\ld}_{max}^{-1} \lt\Vert 1_M \rt\Vert^2 \\
							 & = \bar{\ld}_{max}^{-1} M,
	}which diverges to infinity as $M$ increases. Therefore, $(1_M' \Sigma^{-1} 1_M)^{-1} \le \bar{\ld}_{max}M^{-1} \rightarrow 0 $ as $M \rightarrow \infty$. \qed

\end{appendix}

\bibliographystyle{chicago}
\bibliography{CSA-quantile}

\end{document}